# Revealing complex optical phenomena through vectorial metrics

Chao He[1,*,†], Jintao Chang[2,3,†], Patrick S. Salter[1,†], Yuanxing Shen[3], Ben Dai[4], Pengcheng Li[3], Yihan Jin[1], Samlan Chandran Thodika[5], Mengmeng Li[1], Aziz Tariq[6], Jingyu Wang[1], Jacopo Antonello[1], Yang Dong[3], Ji Qi[7], Jianyu Lin[8], Honghui He[3,*], Daniel S. Elson[8], Min Zhang[9], Hui Ma[2,3,*] and Martin J. Booth[1,*]

[1]*Department of Engineering Science, University of Oxford, Parks Road, Oxford, OX1 3PJ, UK*

[2]*Department of Physics, Tsinghua University, Beijing 100084, China*

[3]*Guangdong Engineering Center of Polarization Imaging and Sensing Technology, Tsinghua Shenzhen International Graduate School, Tsinghua University, Shenzhen 518055, China*

[4]*Department of Statistics, University of Minnesota, Minneapolis, MN 55455, USA*

[5]*University Bordeaux, CNRS, LOMA, UMR 5798, Talence, France*

[6]*Department of Physics, Mirpur University of Science and Technology, Mirpur, AJK-10250, Pakistan*

[7]*Research Center for Intelligent Sensing, Zhejiang Lab, 310000, China*

[8]*Hamlyn Centre for Robotic Surgery, Imperial College London, London SW7 2AZ, UK*

[9]*Respiratory Department, Shenzhen Second People's Hospital, Shenzhen 518055, China*

[†]*These authors contributed equally to this work*

[*]*Corresponding authors:*

chao.he@eng.ox.ac.uk; he.honghui@sz.tsinghua.edu.cn; mahui@tsinghua.edu.cn; martin.booth@eng.ox.ac.uk

**Advances in vectorial polarisation-resolved imaging are bringing new capabilities to applications ranging from fundamental physics through to clinical diagnosis. Imaging polarimetry requires determination of the Mueller matrix (MM) at every point, providing a complete description of an object's vectorial properties. Despite forming a comprehensive representation, the MM does not usually provide easily-interpretable information about the object's internal structure. Certain simpler vectorial metrics are derived from subsets of the MM elements. These metrics permit extraction of signatures that provide direct indicators of hidden optical properties of complex systems, while featuring an intriguing asymmetry about what information can or cannot be inferred via these metrics. We harness such characteristics to reveal the spin-Hall effect of light, infer microscopic structure within laser-written photonic waveguides, and conduct rapid pathological diagnosis through analysis of healthy and cancerous tissue. This provides new insight for the broader usage of such asymmetric inferred vectorial information.**

## Introduction

Light and the optical properties of matter have long been harnessed across different areas of research and application[1]. Rich information can be provided by the vectorial nature of light and its transformation by objects and there is still plenty of opportunity to explore these properties further[2-8]. The way in which an object transforms the vectorial properties of light can be described by the Mueller matrix (MM), which consists of 16 elements ($m_{kl}; k, l = 1,2,3,4$)[9-11]. While the MM forms a comprehensive representation, it does not necessarily provide an intuitive link to the optical properties or structure of the object[9-11]. That is to say, vectorial characteristics of the object, like diattenuation, retardance, and depolarisation (see Fig. 1a), are encoded within the MM elements[9-11].

It has been shown previously that certain symmetries or asymmetries in the MM can be related to physical structures or optical phenomena in the object, when the original targets contain layered optical structures[11]. Useful quantities can be defined, for example, as the difference between diagonally opposite elements of the MM[12, 13]. However, these quantities, have not yet been put forward within a unified structure, thus enabling further intuitive understanding and wider practical use.

Here we first define a universal metric $\Delta$ derived from subsets of elements of the MM, which we refer to here as a "vectorial metric".

$$\Delta(\boldsymbol{u}, \boldsymbol{v}) = \begin{cases} \boldsymbol{u} - \boldsymbol{v} \quad , & d = 1, \\ \|\boldsymbol{u}\|_2 - \|\boldsymbol{v}\|_2, & d \geq 2, \end{cases}$$

where the vectors $(\boldsymbol{u}, \boldsymbol{v})$ consist of MM elements, $\boldsymbol{u} = \left(m_{k_1,l_1}, \cdots, m_{k_d,l_d}\right)'$, and $\boldsymbol{v} = \left(m_{l_1,k_1}, \cdots, m_{l_d,k_d}\right)'$, and $\|\cdot\|_2$ is a vector $l_2$-norm. The general metric encompasses several useful forms, which we will refer to as Metrics 1 to 4 ($\mathrm{M}_1$ to $\mathrm{M}_4$). See Fig. 1b, **Supplementary Notes 1 and 2** for their details.

We point out an asymmetric inference between the metric value and the complexity of objects (Fig. 1c). Such inference is important, as it constrains the physical information that the metrics could extract from objects, such as how multiple layers are constituted (see Fig. 1c, **Supplementary Note 2**). Hence, it not only explains the scope of interpretation of the $\Delta$, but also further applications.

Using this newly unified representation, we then emphasise illustration of the derived vectorial metrics through MM measurements of various objects – objective lenses, photonic waveguides, and biological tissue. Through analysis of such metrics, we reveal the appearance of circular retardance ($M_1$); the combination of linear diattenuation and linear polarisance ($M_2$), multi-layered linear retardance ($M_3$); and the sequence of a linear diattenuator and a linear retarder ($M_4$). Furthermore, practical applications are also enabled: harnessing the spin-Hall effect of light with a GRIN objective lens ($M_1$); conducting vectorial analysis of silica-based photonic waveguides ($M_2$ and $M_3$); and showing rapid cancer diagnosis and layer characterisation of lung tissue ($M_4$).

Moreover, new metrics could be developed that enable extraction of more useful physical information about the target and benefit more applications; while the asymmetric inference behind the metrics can also be explored again (see Discussion). Future impact of these developments could range from quantum physics to clinical diagnosis.

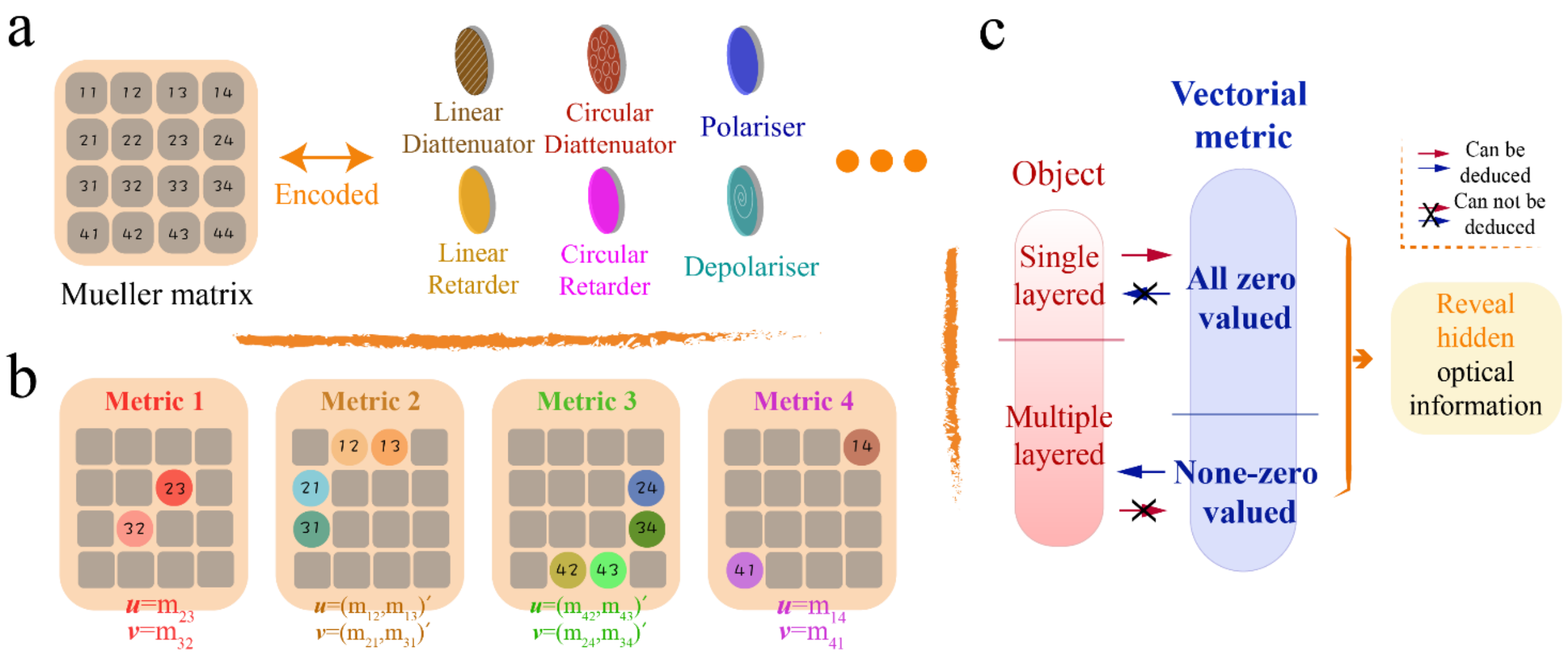

**Figure 1 MM and the asymmetric inference of vectorial metrics.** (a) Different vectorial optical properties that are encoded in a MM. (b) Four vectorial metrics and related elements in their MMs. (c) A summary of the asymmetric inference network of vectorial metrics. The blue and red arrows represent the mathematical inference. Detailed explanations see Supplementary Note 2.

**Revealing spin Hall effect of light**

The spin Hall effect of light (SHEL) has been gaining large attention in fundamental research, such as geometric phase investigations or angular momentum (AM) transfer processes[14-19], and in applications such as optical edge detection for characterising boundaries[15]. Traditional methods to validate SHEL include interferometry[16] or quantum weak measurement[17, 18]; however they are incapable of measuring or tracing SHEL's presence in the multi-layered optical phenomena that result in a complex inhomogeneously polarised beam. Hence, MM analysis has become a powerful tool to deal specifically with such complex scenarios[19]. Here we demonstrate the use of $M_1$ to validate the presence of a circular birefringence (CB) gradient, which serves as an efficient way to trace SHEL's presence in complex systems[18]. This is illustrated through analysis of the optical properties of a graded index (GRIN) lens[20] under the condition of oblique illumination.

GRIN lenses are widely used in compact imaging systems[20]. Their intrinsic radially symmetric linear birefringence, which can also be considered as a systematic polarisation aberration, has been studied before[20]. However, the performance of such birefringence with obliquely incident light has not yet been investigated; such a configuration is relevant, for example, to scanning laser microscopy through a GRIN lens endoscope[21]. Under such conditions, the beam would no longer obey the meridional approximation with respect to the gradually changing refractive index of the GRIN lens[20, 22], such that the effective aberration follows a complex development process rather than being modelled by a pure linear retarder assembly. In Fig. 2a we show two cases of obliquely illuminated GRIN endoscope lenses, with two different incident angles of 5° and 7°. In Fig. 2b and 2c, the corresponding MMs, output vector beams, the equivalent retarder fast axis orientations, a CB gradient example and $M_1$ are given (see **Supplementary Note 3** for the Mueller-Stokes measurement approach and other decomposition methods). A plot of the value of $M_1$ with changing illumination angle is shown in Fig. 2d (see details in **Supplementary Notes 2 and 4**).

Additionally, the fast axis distributions of the two exemplar cases (Fig. 2b) demonstrate the transformation of intrinsic topological charges[19] to support the existence of SHEL from another perspective – that of AM (see **Supplementary Note 4** for a detailed analysis). This demonstration newly reveals the AM transfer process in a GRIN lens under the non-meridional plane geometry. As the GRIN lens is also an imaging device, it has the potential to be used as a novel imaging SHEL device, which presents its own applications such as focus manipulation[23]. As the illumination angle of the light beam relative to the lens can be controlled, the SHEL in such a system is field-controllable[19], which would give additional degrees of freedom to harness the SHEL through such a device.

Here, for the first time, by analysing the hidden information in MM vector images of an oblique illuminated GRIN lens system through $M_1$, we validated the presence of the CB gradient in such system, hence confirming the existence of SHEL.

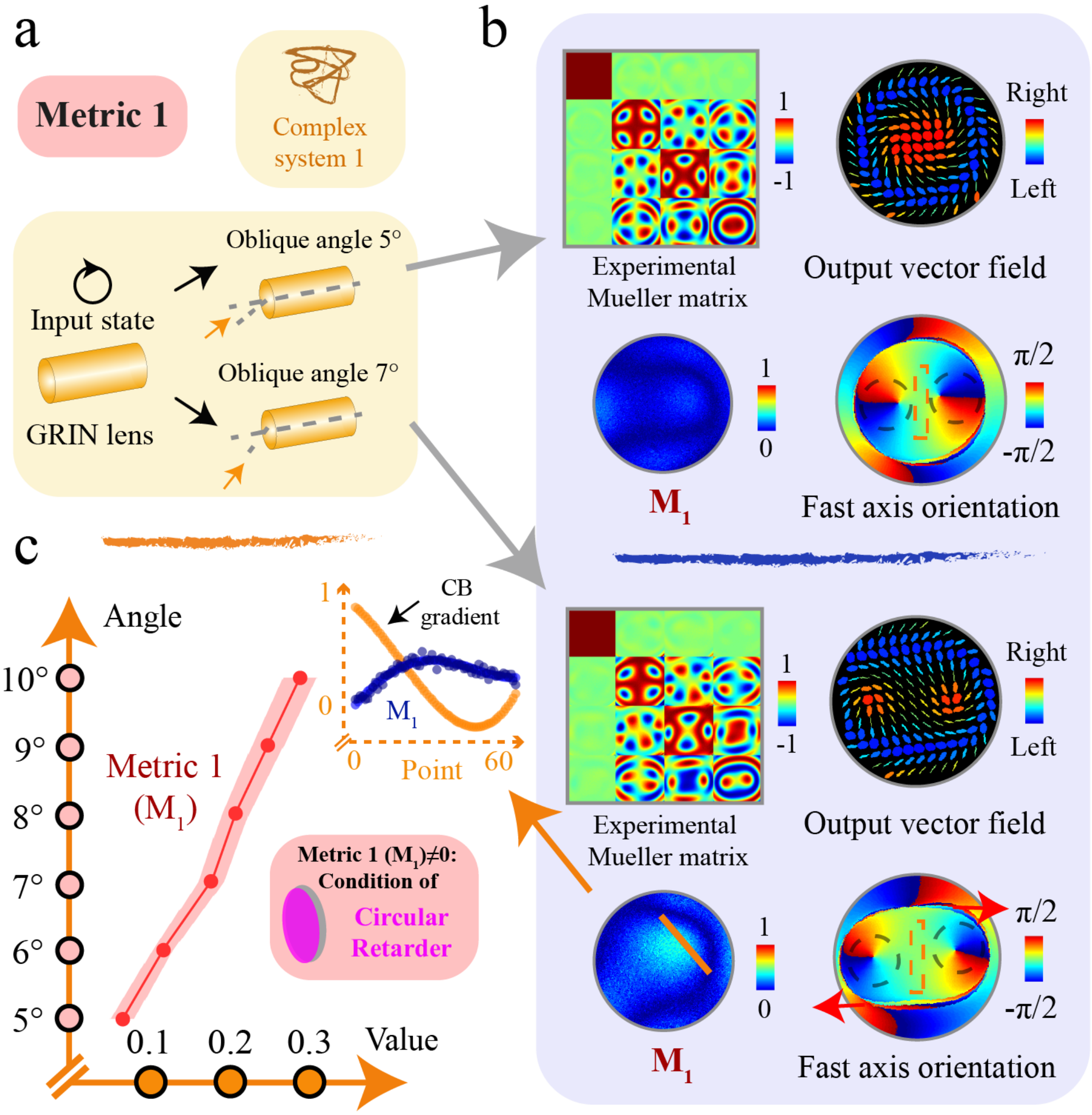

**Figure 2 GRIN lens with its decoupled vectorial information.** (a) GRIN lens with right hand circular polarised light input, under oblique incident angle at 5° and 7° respectively. (b) Their related experimental MMs, output vector fields, $M_1$, fast axis orientations, as well as an example of a plot of CB gradient sampled via the yellow solid line within sub-figure of $M_1$. (c) Value of $M_1$ under different oblique incident angles of this system (we use absolute value in this section for simplicity), from 5° to 10° with 1° interval. The value is extracted via the middle linear gradient area (orange dotted rectangle, detailed in Supplementary Note 3). The shaded area represents the standard deviation. Details of the relationship between $M_1$ and CB can be found in Supplementary Note 2.

**Revealing vectorial phenomena in laser written waveguides**

We have seen above how the polarisation aberration of a GRIN lens leads to mechanical effects that have optical consequences that are revealed through the vectorial metrics. Similar effects can occur during the fabrication of other photonic structures. We show here how the vectorial metrics $M_2$ and $M_3$ can be used to extract useful information about laser written waveguides[24] from complex MM measurements.

Photonic integrated circuits (PICs), comprising a 3D network of waveguides embedded in a glass substrate, are used in a range of applications and may be written by an ultrashort pulsed laser[25]. These so-called direct laser written PICs have found particular application in photon-based quantum information processing[26-28]. In such applications, two waveguide parameters are of critical importance: low loss and low polarization dependence[29]. Previous studies of these two parameters have focused on characterising the properties of the output light beam, such as output intensity under certain incident states of polarisation (SOPs), or the output SOP measured by point Stokes vector polarimetry[29]. Those measurements and further characterisation have intrinsic limitations, since they cannot comprehensively represent the full vectorial properties[9-11] of the PICs. The subsequent optimisation processes for these waveguides are therefore limited since the light matter interaction has not been fully considered.

By analysing the MM vector images of laser-written PICs in fused silica (Fig. 3a), we take a new perspective to reveal their intrinsic optical properties. This analysis first focuses on the value of $M_2$ and $M_3$. $M_2$ validates the existence condition of linear polarisance and diattenuation properties; $M_3$ reveals the condition of multi-layered linear retarders (see **Supplementary Notes 1 and 2**).

In Fig. 3b, the experimental MMs are given for two straight waveguides fabricated at different laser writing pulse energies of 42 nJ and 67 nJ. Further manufacturing details are provided in **Supplementary Note 5**. Selected optical properties for the waveguides from the MM imaging are plotted in Fig. 3b, as well as quantification in Fig. 3c as a function of the fabrication pulse energy. The existence of non-zero $M_2$ indicates that such a system exhibits stronger diattenuation and less polarisance[30]; this is equivalent to a layered structure consisting of a diattenuator followed by a depolariser (see **Supplementary Note 6**). Such properties can be a result of the combination of Rayleigh scattering[31] and complex nano-grating[32, 33] induced anisotropic scattering[34-36] inside the photonic waveguide. It is known that with increasing energy of the writing laser pulses, the status of such nano-structures can be changed during the fabrication process[37], here correlating with a reduction in $M_2$ and increasing depolarisation. We validated the possibilities via Monte Carlo simulation with respect to the changes of the anisotropic scattering. Such simulation has been widely employed for quantitative analysis of the interactions between polarized photons and complex media, especially in the presence of multiple scattering, where analytical solutions are not available[34, 35] (see **Supplementary Note 6**). We also observed the existence of $M_3$ in the same waveguides shown in Fig. 3b. We can see that asymmetry occurs between the fourth row and the fourth column of the MM, which supports the existence of multi-layered linear retarders featuring directional dependency. Such an effect is also detailed in **Supplementary Note 6**, through measurement of different retardance for waveguides written with opposite scanning direction, a further validation of the quill effect in laser fabrication[38].

The observation of trends in the vectorial metric values and their analysis give us a better insight into the vectorial properties of the photonic waveguides, including better understanding of the loss and polarisation effects as well as providing guidance for further customized fabrication. It reveals that, 1) different types of complex layered polarisation structures exist inside the waveguides (validated via $M_2$ and $M_3$); as $M_2$ can be interpreted as indicating an anisotropic absorption layer before a depolarisation layer, $M_3$ can be used to characterize the contributions from stress birefringence induced, form birefringence induced, as well as scattering induced retardance; 2) $M_2$ decreased with increasing pulse energy, while the scattering induced depolarisation increased. These observations imply that under such a writing parameter regime, the balance of polarisation dependence loss ($M_2$) and scattering level (depolarisation) should be considered while choosing the pulse energy to optimise waveguides for different applications (e.g., 67 nJ would have less polarisation dependence loss but would show relatively higher scattering). Such observations can only be made via full analysis of vectorial properties.

For the first time, by harnessing the vectorial metrics of MM images, we have performed vectorial analysis of the optical properties of the direct laser written waveguides. These insights will help optimise fabrication processes to improve the properties of advanced PICs. They also provide a new perspective for investigation of light matter interaction in the waveguides due to fabrication effects.

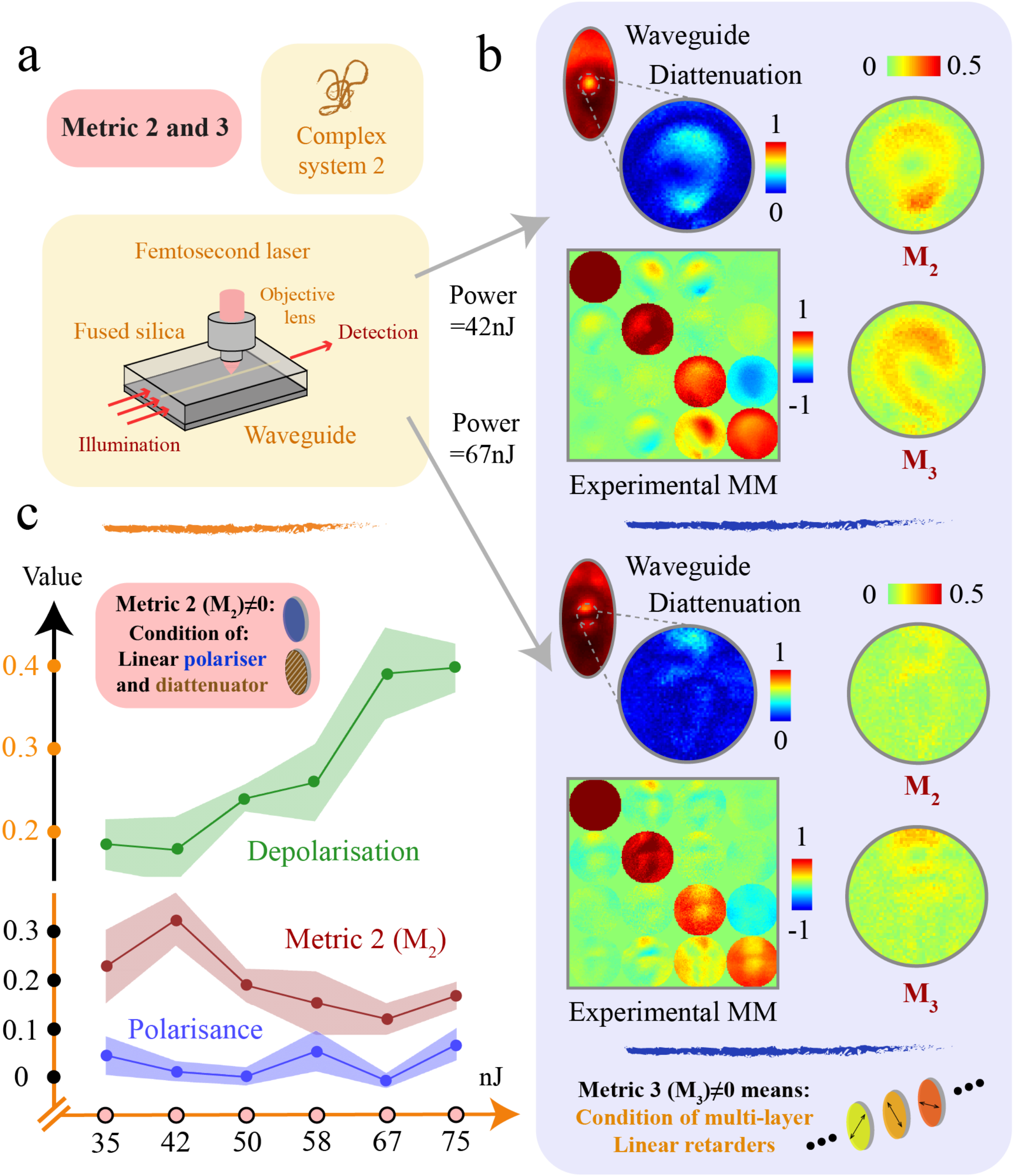

**Figure 3 Direct laser written waveguides with their MMs and vectorial metric analysis** (a) A sketch of the geometry of the direct laser writing process and the illumination and detection paths of the subsequent imaging process. (b) Example experimental MMs, value of metrics for waveguides written with laser pulse energies of 42nJ and 67nJ, respectively. (c) The value of depolarisation, $M_2$ and polarisance of different waveguides with respect to writing pulse energy. The statistical analysis was performed on the chosen sub-regions (detailed in Supplementary Note 5). The shaded area represents the standard deviation. The relationship between $M_2$ (and $M_3$) to polarisation properties can be found in Supplementary Note 2, 6 and 8.

**Revealing structural information of cancerous tissues for differentiation**

Cancer is one of the biggest threats to human health. Clinically, non-small cell lung carcinoma is the primary form of lung cancer constituting about 85% of all cases[39, 40]. There have been numerous proposed scalar and vectorial optical methods, based upon intensity imaging, for differentiating the normal and cancerous tissues, as well as exploring their related pathological microstructures[41-43]. Vectorial analysis of biomedical tissue has a shorter history but is continuously gaining more attention[41-43].

Here, for the first time, by analysing the vectorial metrics derived from MM images of normal human lung tissues and non-small cell lung carcinoma tissues, we differentiate the tissue types via the lateral images while revealing their pathological structural information using axial information via metric $M_4$. Note that $M_4$ also represents the sequence of layers of linear retardance and linear diattenuation (see **Supplementary Note 1**).

We measured the MMs of five pairs of normal lung tissue areas (H1 to H5) and human non-small cell lung carcinoma (C1 to C5) tissue areas from sample 1-5 (each sample contains regions of H and C, decided by pathologists from Shenzhen Second People's Hospital). Figure 4a shows the sketch of normal lung tissue and its alveoli, as well as the alveoli with pulmonary fibrosis processes[44]. For each sample, we selected one normal-tissue-based region and one cancerous-tissue-based region as demonstrated in Fig. 4a. For the measured MMs, absolute values of $M_4$ and decomposed retardance, which has been used as a quantitative parameter for cancer detection[20, 45] (see **Supplementary Notes 3** and **7**), as well as other parameters of two randomly chosen regions are illustrated in Fig. 4b. For each region we sampled ten points and extracted the value of two parameters for later statistical analysis (for methods refer to Ref[20]). The statistical histogram of the data from sample 1 is shown in Fig. 4c. It can be seen that the value of $M_4$ for all cancerous areas are significantly larger in magnitude than those in normal areas, and it shows similar statistical variation as the retardance. Though decomposed retardance has previously been proposed as a useful measure for abnormal fibrous tissue areas, the $M_4$ has more advantages: it provides a physically interpretable, simply calculated index for quick clinical diagnosis and cancer boundary detection. Details of all original data, statistical analysis including P-value analysis, as well as sampling and processing methods can be found in **Supplementary Note 7**.

Furthermore, as mentioned earlier, the existence of $M_4$ also relates to the sequential structure along the optical axis of a multi-layered system. Here the determined relationship based $M_4$ for those human lung carcinoma tissue pairs is shown in Fig. 4d (see **Supplementary Note 1** for details). It effectively indicates the possibility that within some regions of lung cancer tissue, there exists a layered structure which may be treated as a linear retarder followed by a linear diattenuator. Such a fibrous structure model is supported with a detailed Monte Carlo simulation approach[34, 35] (see **Supplementary Note 8** for details) and can potentially be used as evidence in support of the corresponding morphological explanations[46]. Such parameters (related to newly revealed fibrous structures) are likely to be useful in quick, label-free determination of the specific cancerous development stages or micro-structural identification. Validation of this will require further research in the fields of both polarisation biophotonics and clinical pathological diagnosis.

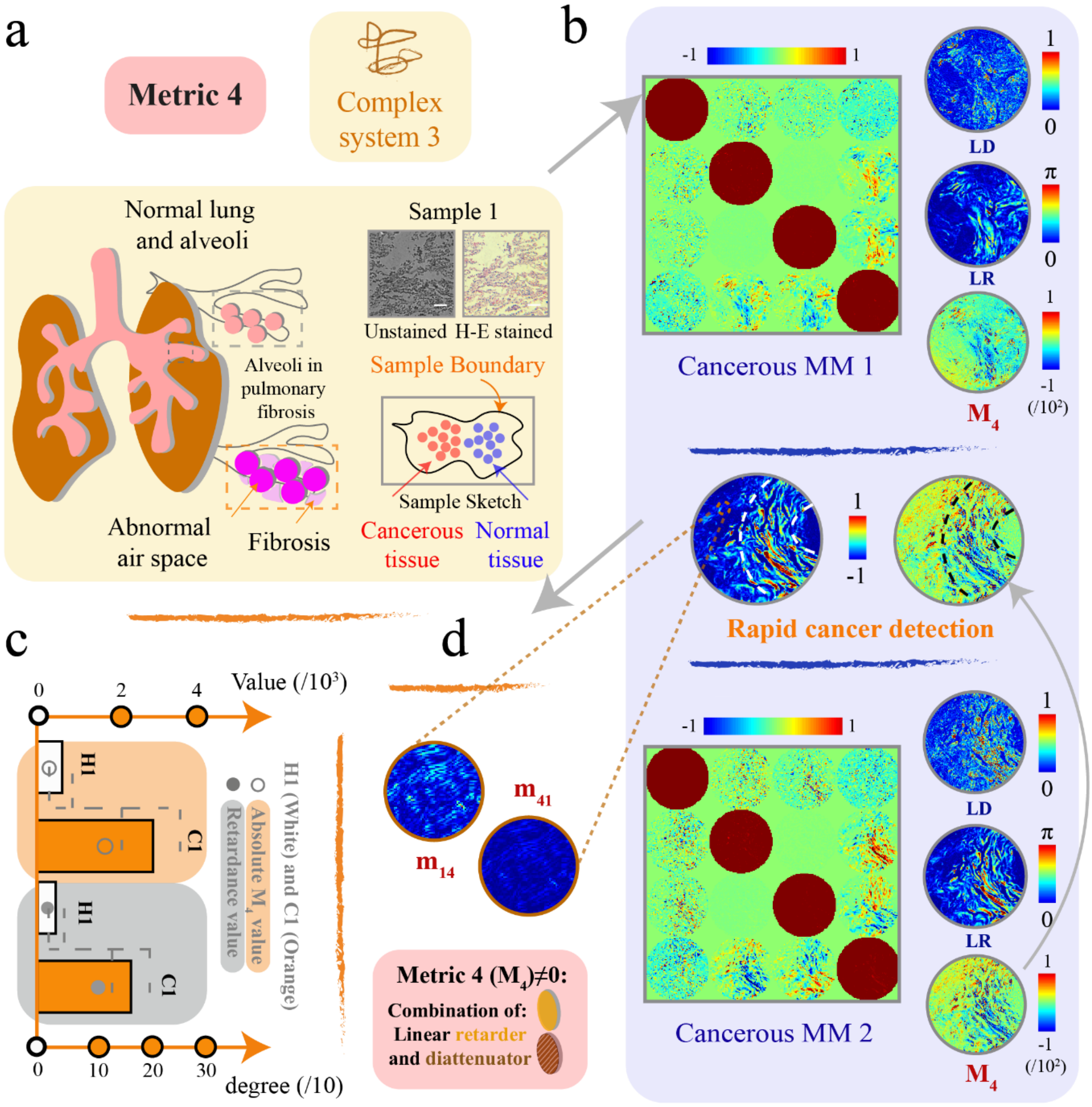

**Figure 4 Normal/cancerous lung tissue samples with their MMs and metric information** (a) Sketches of normal lung tissue and alveoli and abnormal lung tissue with fibrosis. Demonstration for samples 1 (unstained sample and its H-E stained counterpart), showing a sketch with corresponding random sampling points in both cancerous and normal areas (see method in Supplementary Note 7). Scale bar: 50 µm. (b) Experimental MMs sampled from sample C1, alongside with value of linear diattenuation (LD), linear retardance (LR), and $M_4$ at two randomly chosen regions. Note the scale used for LD and LR (and related MM elements) has been amplified by a factor of 100 for better visualisation. (c) Demonstration of the data from samples 1 (H1 and C1 parts); the bar chart shows mean value and the standard deviation of $M_4$ and retardance. Data were decomposed from the MMs for 10 points per each region of the sample. (d) A selected small region of $m_{14}$ and $m_{41}$ of the second cancerous MM is shown as well. This indicates the structure may have double-layered format – a sequence of linear retarder and linear diattenuator; details can be found in Supplementary Notes 1, 2 and 7.

## Discussion

In this work, we took advantage of the vectorial metrics derived from the MM – that can provide a simple representation of complex vectorial phenomena – to demonstrate SHEL, to analyse the vectorial characteristics of laser written waveguides, and to facilitate the discrimination between healthy and cancerous tissue. It has shown that simplified metrics based upon simple combinations of MM elements can reveal useful information about specimen properties across diverse areas of technology. Further insight could again be derived from extension of the use of these metrics into other areas (Metric 1-4; or alternative metrics, see more in **Supplementary Note 9**). Furthermore, it is informative to consider the relationships between objects, vectorial metrics and the complete MMs – in particular the inferences that can be drawn between each of them.

Some complex multiple layered structures can possess MMs with properties in common with certain simple, single layered structures. These properties can be revealed in a straightforward way through the use of appropriately defined vectorial metrics. Hence the metrics can provide an intermediate way to link the properties of objects with the intrinsic characteristics of a MM. Based on this observation and the asymmetric inference that is used in the main article, more complex connections can be revealed (see Fig. 5a), based on three spaces: A (representing objects), B (representing the vectorial metrics), and C (representing the MMs). They are separate but linked concepts, between which there exists a complex inference relationship. The inference relationships between A and B have been explained and applied in the content above, where we characterised its asymmetry. A similar asymmetric inference can be found between the spaces B and C (see Fig. 5a), hence forming an overall complex connection via space B. The MM space C can also be divided into two sub-spaces $C_1$ and $C_2$. $C_1$ represents MMs that demonstrate a connection between seemingly disparate physical systems. For example, this could mean that a single layered system ($A_1$) and a multi-layered system ($A_2$) can exhibit same forms of MM, while their internal structures are considerably different (see Fig. 5b (i) for the concept; and 5b (ii) for a real example consisting of a GRIN lens). The latter sub space $C_2$ represents those MMs that do not demonstrate such a connection between $A_1$ and $A_2$. Such property is linked via space B and hence can be revealed using vectorial metrics. Understanding such entangled phenomena could help enable various new potential applications, such as providing an advanced passive vectorial aberration compensation strategy.

We demonstrate here a potential new metric ($M_5$) that working under such scope. For instance, we first use Monte Carlo simulation approach[34, 35] to simulate a multi-layered complex birefringence system that includes the aberration, then use the value of $M_5$ to serve as a feedback control. A general process would include three steps: 1) run the simulation and check the value of $M_5$ simultaneously; 2) record the multi-layered system's parameters when $M_5$ appeared to be the expected value; 3) adopt such generated multi-layered configuration in reality. Take a GRIN lens system in a specific configuration as an example (Fig. 5b (ii)). Here, the complex polarisation aberration of the GRIN lens endoscope system would be equivalent to a simplified system – a retarder with uniform fast axis orientation, which also exhibits the same zero metric. This would be clearly seen if using metric $M_5$. Hence, further aberration compensation would only require a phase correction device[47] if used with x- or y -polarised illumination (Fig. 5b (iii)). This case shows that the vectorial metric not only maintains the ability for characterisation of complex optical properties, but can also be used as a simple measure for feedback in an optimisation procedure – see Fig. 5c; and **Supplementary Note 9** for details.

These observations indicate that a large unexplored space exists for research into related vector MM images. Our work paves the way for the broader usage of such entangled vectorial information, which is inherently connected through the vectorial metrics, in further applications spanning from quantum physics to clinical applications.

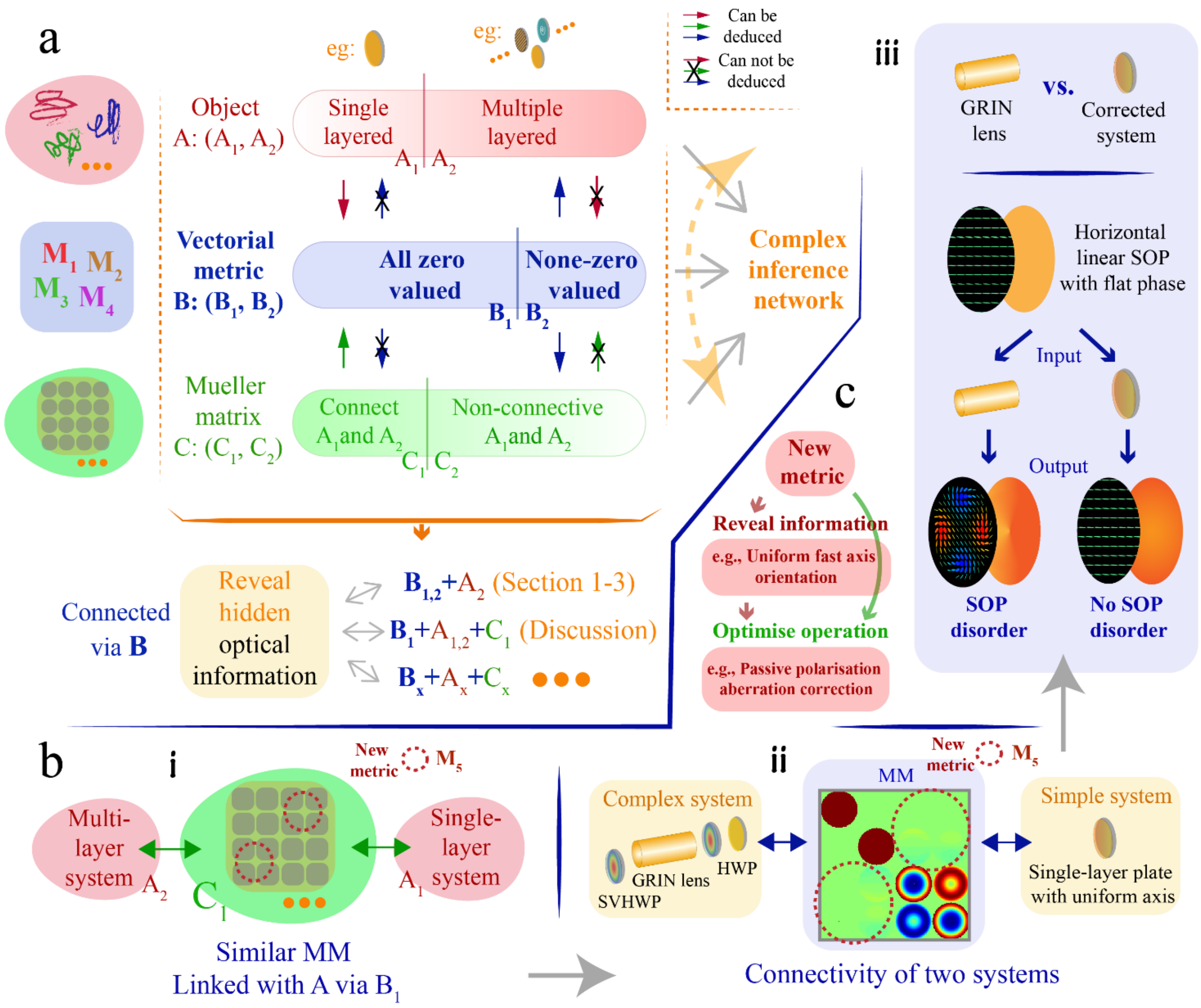

**Figure 5 Relationships between the three spaces and passive polarisation aberration compensation.** (a) The complex inference network among object (space A), vectorial metrics (space B) and MM (space C). (b) (i) and (ii) show a sketch of the connectivity between A, B and C, potential new metric ($M_5$; defined with two diagonally opposed 2x2 blocks; details see Supplementary Note 9), as well as a demonstration of passive polarisation compensation using GRIN lens and spatial half waveplate array (SVHWP). HWP: half wave plate. (iii) An illustration of the effect of aberration compensation. (c) Such spaces are linked with new metric. The metric can first reveal hidden physical information of a complex system, such as uniform axis orientation in this application, then can optimise the operations, such as achieving an aberration compensated system.

## Acknowledgments

The project was supported by the European Research Council (AdOMiS, no. 695140) (C.H. and M.J.B.); Engineering and Physical Sciences Research Council (UK) (EP/R004803/01) (P.S.S.); National Natural Science Foundation of China (11974206, 61527826) (H.M.); Shenzhen Fundamental Research and Discipline Layout Project (No. JCYJ20170412170814624) (H.H., M.Z. and H.M.); H2020-MSCAIF-2018 Programme, Grant No. 838199 (S.C.T.). This work was approved by the Ethics Committee of the Shenzhen Second People's Hospital. The authors thanks Meiquan Xu from Shenzhen Second People's Hospital (pathology department) for pathological assistance and advice. The first author C.H. would like to thank Prof Nirmal K Viswanathan from the University of Hyderabad, Dr Jun Guan from University of Oxford, Dr Mohan Wang from University of Oxford, Prof Yonghong He from Tsinghua University for discussions about SHEL, glass-ceramics, nano-grating, and vector information analysis.

## Competing interests

The authors declare no competing interests.

## Author contributions

C.H., J.C., H.H. and M.J.B. conceived the project. J.C., P.L., J.A., S.C.T. and C.H. contributed to the SHEL section. P.S.S. and C.H. contributed to the waveguide section; P.S.S. wrote the waveguides. H.H., Y.S., Y.D., C.H., D.S.E. and H.M. contributed to the biomedical sample section; M.Z. prepared the tissue samples. H.H., Y.S. and H.M. contributed to the MC modelling. C.H. and M.J.B. contributed to the aberration compensation section. Y.J., J.C., M.L., J.W., J.A., Y.S., J.Q., J.L., C.H. and P.S.S. contributed to data processing and data analysis. B.D., A.T., C.H., P.L., D.S.E. and H.M. contributed to the unified metrics. C.H., J.C., H.H., B.D. and M.J.B. contributed to the complex inference network. C.H. prepared the figures. C.H. and M.J.B. wrote the paper. All authors participated in editing of the paper.

## Additional Information

Correspondence and request for materials should be addressed to C.H., H.H., H.M. or M.J.B.

**References**

1. Born, M., Wolf, E. *Principles of optics: electromagnetic theory of propagation, interference and diffraction of light* (Elsevier, 2013)

2. Bliokh, K. Y., Rodríguez-Fortuño, F. J., Nori, F., Zayats, A. V. Spin–orbit interactions of light. *Nat. Photonics*. **9**, 796-808 (2015).

3. Forbes, A., De Oliveira, M., Dennis, M. R. Structured light. *Nat. Photonics*. **15**, 253-262 (2021).

4. Wang, J., Sciarrino, F., Laing, A., Thompson, M. G. Integrated photonic quantum technologies. *Nat. Photonics*. **14**, 273-284 (2020).

5. Dholakia, K., Drinkwater, B. W., Ritsch-Marte, M. Comparing acoustic and optical forces for biomedical research. *Nat. Rev. Phys*. **2**, 480-491 (2020).

6. Dorrah, A. H., Rubin, N. A., Zaidi, A., Tamagnone, M., Capasso, F. Metasurface optics for on-demand polarization transformations along the optical path. *Nat. Photonics*. **15**, 287-296 (2021).

7. Hafi, N., Grunwald, M., Van Den Heuvel, L. S., Aspelmeier, T., Chen, J.-H., Zagrebelsky, M., *et al.* Fluorescence nanoscopy by polarization modulation and polarization angle narrowing. *Nat. Methods*. **11**, 579 (2014).

8. Jin, D., Xi, P., Wang, B., Zhang, L., Enderlein, J., Van Oijen, A. M. Nanoparticles for super-resolution microscopy and single-molecule tracking. *Nat. Methods*. **15**, 415-423 (2018).

9. Goldstein, D. H. *Polarized light Book Polarized light* (CRC press, Boca Raton, 2017)

10. Chipman, R. A., Lam, W.-S. T., Young, G. *Polarized Light and Optical Systems.* (CRC press, Boca Raton, 2018).

11. Pérez, J. J. G., Ossikovski, R. *Polarized light and the Mueller matrix approach Book Polarized light and the Mueller matrix approach* (CRC press, Boca Raton, 2017)

12. Li, P., Lv, D., He, H., Ma, H. Separating azimuthal orientation dependence in polarization measurements of anisotropic media. *Opt. Express*. **26**, 3791 (2018).

13. Li, P., Tariq, A., He, H., Ma, H. Characteristic Mueller matrices for direct assessment of the breaking of symmetries. *Opt. Lett*. **45**, 706-709 (2020).

14. Ling, X., Zhou, X., Huang, K., Liu, Y., Qiu, C.-W., Luo, H., *et al.* Recent advances in the spin Hall effect of light. *Rep. Prog. Phys*. **80**, 066401 (2017).

15. Zhou, J., Qian, H., Chen, C.-F., Zhao, J., Li, G., Wu, Q., *et al.* Optical edge detection based on high-efficiency dielectric metasurface. *PNAS*. **116**, 11137-11140 (2019).

16. Prajapati, C., Pidishety, S., Viswanathan, N. K. Polarimetric measurement method to calculate optical beam shifts. *Opt. Lett*. **39**, 4388-4391 (2014).

17. Hosten, O., Kwiat, P. Observation of the Spin Hall Effect of Light via Weak Measurements. *Science*. **319**, 787-790 (2008).

18. Bliokh, K. Y., Samlan, C. T., Prajapati, C., Puentes, G., Viswanathan, N. K., Nori, F. Spin-Hall effect and circular birefringence of a uniaxial crystal plate. *Optica*. **3**, 1039 (2016).

19. Samlan, C. T., Viswanathan, N. K. Field-controllable Spin-Hall Effect of Light in Optical Crystals: A Conoscopic Mueller Matrix Analysis. *Sci. Rep*. **8** (2018).

20. He, C., Chang, J., Hu, Q., Wang, J., Antonello, J., He, H., *et al.* Complex vectorial optics through gradient index lens cascades. *Nat. Commun*. **10**, 1-8 (2019).

21. Huland, D. M., Brown, C. M., Howard, S. S., Ouzounov, D. G., Pavlova, I., Wang, K., *et al.* In vivo imaging of unstained tissues using long gradient index lens multiphoton endoscopic systems. *Biomed. Opt. Express*. **3**, 1077-1085 (2012).

22. Ghatak, A. K. *Optics* (McGraw-Hill, City of New York, 2005)

23. Zhou, J., Qian, H., Hu, G., Luo, H., Wen, S., Liu, Z. Broadband Photonic Spin Hall Meta-Lens. *ACS Nano*. **12**, 82-88 (2018).

24. Salter, P. S., Booth, M. J. Adaptive optics in laser processing. *Light Sci. Appl*. **8** (2019).

25. Della Valle, G., Osellame, R., Laporta, P. Micromachining of photonic devices by femtosecond laser pulses. *JOptA*. **11**, 013001 (2008).

26. Crespi, A., Osellame, R., Ramponi, R., Brod, D. J., Galvão, E. F., Spagnolo, N., *et al.* Integrated multimode interferometers with arbitrary designs for photonic boson sampling. *Nat. Photonics*. **7**, 545-549 (2013).

27. Marshall, G. D., Politi, A., Matthews, J. C. F., Dekker, P., Ams, M., Withford, M. J.*, et al.* Laser written waveguide photonic quantum circuits. *Opt. Express*. **17**, 12546-12554 (2009).

28. Tillmann, M., Dakić, B., Heilmann, R., Nolte, S., Szameit, A., Walther, P. Experimental boson sampling. *Nat. Photonics*. **7**, 540-544 (2013).

29. Guan, J., Liu, X., Salter, P. S., Booth, M. J. Hybrid laser written waveguides in fused silica for low loss and polarization independence. *Opt. Express*. **25**, 4845-4859 (2017).

30. Chipman, R. A. Depolarization index and the average degree of polarization. *Appl. Opt*. **44**, 2490-2495 (2005).

31. Eaton, S. M., Chen, W.-J., Zhang, H., Iyer, R., Li, J., Ng, M. L.*, et al.* Spectral loss characterization of femtosecond laser written waveguides in glass with application to demultiplexing of 1300 and 1550 nm wavelengths. *J. Light. Technol*. **27**, 1079-1085 (2009).

32. Shimotsuma, Y., Kazansky, P. G., Qiu, J., Hirao, K. Self-Organized Nanogratings in Glass Irradiated by Ultrashort Light Pulses. *Phys. Rev. Lett*. **91** (2003).

33. Hirao, K., Miura, K. Writing waveguides and gratings in silica and related materials by a femtosecond laser. *J Non Crys Solids*. **239**, 91-95 (1998).

34. Wang, L., Jacques, S. L., Zheng, L. MCML—Monte Carlo modeling of light transport in multi-layered tissues. *Comput. Methods Programs Biomed*. **47**, 131-146 (1995).

35. Yun, T., Zeng, N., Li, W., Li, D., Jiang, X., Ma, H. Monte Carlo simulation of polarized photon scattering in anisotropic media. *Opt. Express*. **17**, 16590 (2009).

36. He, H., Liao, R., Zeng, N., Li, P., Chen, Z., Liu, X.*, et al.* Mueller Matrix Polarimetry—An Emerging New Tool for Characterizing the Microstructural Feature of Complex Biological Specimen. *J. Light. Technol*. **37**, 2534-2548 (2019).

37. Hnatovsky, C., Taylor, R. S., Rajeev, P. P., Simova, E., Bhardwaj, V. R., Rayner, D. M.*, et al.* Pulse duration dependence of femtosecond-laser-fabricated nanogratings in fused silica. *Appl. Phys. Lett*. **87**, 014104 (2005).

38. Kazansky, P. G., Yang, W., Bricchi, E., Bovatsek, J., Arai, A., Shimotsuma, Y.*, et al.* “Quill” writing with ultrashort light pulses in transparent materials. *Appl. Phys. Lett*. **90**, 151120 (2007).

39. Herbst, R. S., Morgensztern, D., Boshoff, C. The biology and management of non-small cell lung cancer. *Nature*. **553**, 446-454 (2018).

40. Reck, M., Heigener, D. F., Mok, T., Soria, J.-C., Rabe, K. F. Management of non-small-cell lung cancer: recent developments. *The Lancet*. **382**, 709-719 (2013).

41. Ghosh, N., Vitkin, A. I. Tissue polarimetry: concepts, challenges, applications, and outlook. *J. Biomed. Opt*. **16**, 110801 (2011).

42. Tuchin, V. V. Polarized light interaction with tissues. *J. Biomed. Opt*. **21**, 071114 (2016).

43. Ramella-Roman, J., Saytashev, I., Piccini, M. A review of polarization-based imaging technologies for clinical and pre-clinical applications. *JOpt.* **22**, 123001 (2020).

44. Litin, S. C., Nanda, S. *Mayo clinic family health book* (Time Incorporated Home Entertainment, Des Moines, 2009)

45. Dong, Y., Qi, J., He, H., He, C., Liu, S., Wu, J*., et al.* Quantitatively characterizing the microstructural features of breast ductal carcinoma tissues in different progression stages by Mueller matrix microscope. *Biomed. Opt. Express*. **8**, 3643-3655 (2017).

46. King, T. E., Pardo, A., Selman, M. Idiopathic pulmonary fibrosis. *The Lancet*. **378**, 1949-1961 (2011).

47. Booth, M. J. Adaptive optical microscopy: the ongoing quest for a perfect image. *Light. Sci. Appl*. **3**, 165 (2014).

# Revealing complex optical phenomena through vectorial metrics

## – Supplementary information –

He et al.

# Supplementary Note

Chao He[1,*,†], Jintao Chang[2,3,†], Patrick S. Salter[1,†], Yuanxing Shen[3], Ben Dai[4], Pengcheng Li[3], Yihan Jin[1], Samlan Chandran Thodika[5], Mengmeng Li[1], Aziz Tariq[6], Jingyu Wang[1], Jacopo Antonello[1], Yang Dong[3], Ji Qi[7], Jianyu Lin[8], Honghui He[3,*], Daniel S. Elson[8], Min Zhang[9], Hui Ma[2,3,*] and Martin J. Booth[1,*]

[1]*Department of Engineering Science, University of Oxford, Parks Road, Oxford, OX1 3PJ, UK*

[2]*Department of Physics, Tsinghua University, Beijing 100084, China*

[3]*Guangdong Engineering Center of Polarization Imaging and Sensing Technology, Tsinghua Shenzhen International Graduate School, Tsinghua University, Shenzhen 518055, China*

[4]*Department of Statistics, University of Minnesota, Minneapolis, MN 55455, USA*

[5]*University Bordeaux, CNRS, LOMA, UMR 5798, Talence, France*

[6]*Department of Physics, Mirpur University of Science and Technology, Mirpur, AJK-10250, Pakistan*

[7]*Research Center for Intelligent Sensing, Zhejiang Lab, 310000, China*

[8]*Hamlyn Centre for Robotic Surgery, Imperial College London, London SW7 2AZ, UK*

[9]*Respiratory Department, Shenzhen Second People's Hospital, Shenzhen 518055, China*

[†]*These authors contributed equally to this work*

[*]*Corresponding authors:*

chao.he@eng.ox.ac.uk; he.honghui@sz.tsinghua.edu.cn; mahui@tsinghua.edu.cn; martin.booth@eng.ox.ac.uk

## Supplementary Note 1: Mueller matrix and the related polarisation properties.

Either a Jones matrix (JM) or a Mueller matrix (MM) can be used to describe the vectorial optical properties of an object [1-3]. While the JM contains the information of absolute phase, it cannot represent depolarising effects of objects. The MM however can represent depolarisation, but not the overall phase[1-3]. Hence, the MM is used if comprehensive polarisation properties need to be considered. The MM is a 4 by 4 matrix, so consists of 16 elements ($m_{kl}; k, l = 1,2,3,4$). Among those elements, $m_{11}$ represents the change in intensity, and the other 15 elements encode the vectorial properties of the object[1-3]. The relationship between the physical quantities like retardance and these separate MM elements is often ambiguous[1-3].

Numerous decomposition methods have been put forward [4-6] to extract different polarisation parameters from the MM. These include linear/circular polarisance, linear/circular diattenuation, linear/circular retardance, and linear/circular depolarisation[4-6]. The various widely-used decomposition methods include: MM polar decomposition[4], MM transformation parameters[5], and MM anisotropic coefficients[6]. However, those approaches are based on different assumptions such as matrix reciprocity[4], therefore they suffer intrinsic limitations in revealing the relationship to the real physical structure of a specimen – the extracted information is inherently biased to the assumed mathematical structure (some methods are shown in Supplementary Note 3). In this work, we focus on the vectorial information decomposed from the MM that relate to the determined physical information. We refer to these quantities as "vectorial metrics", as further detailed in Supplementary Note 2.

Supplementary Figure 1 shows a general MM and demonstrates several encoded polarisation properties. Three typical MMs are illustrated here: an arbitrary wave plate (described by parameters of retardance and axis orientation), an arbitrary polariser (described by parameters diattenuation and axis orientation), and an arbitrary depolariser (described by the parameter of depolarisation). These are shown in Ref[3]. For a waveplate, the parameter set $(\delta, \theta, \varepsilon)$ represents the retardance value $\delta$, fast axis orientation $\theta$, and the latitude parameter $\varepsilon$ $(-\pi/2 \leq \varepsilon \leq \pi/2)$ that determines the shape of the ellipse. In a polariser, the parameter set $(\theta, \varepsilon)$ encodes the transmission of a state of polarisation (SOP) with the major axis of the ellipse oriented at $\theta$ located at latitude $\varepsilon$ on the Poincaré sphere, where $-\pi/2 \leq \varepsilon \leq \pi/2$. In a depolariser, there exist six degrees of freedom; they are $-1 \leq a, b, c, d, e, f \leq 1$, and the corresponding MM is in a symmetric format. Note that such comprehensive polarisation properties can only be extracted via a MM (or from a JM when the medium exhibits no depolarisation), while such full properties are not accessible by other optical measurement approaches[3].

$$M_{Final} = M_n \dots M_{m+1} \cdot M_m \dots M_2 \cdot M_1, \tag{1-1}$$

Eq. (1-1) models the mixed polarisation properties of a complex system $M_{Final}$, where the system is assumed to have $n$ layers, and $M_m$ represents an arbitrary layer inside such system.

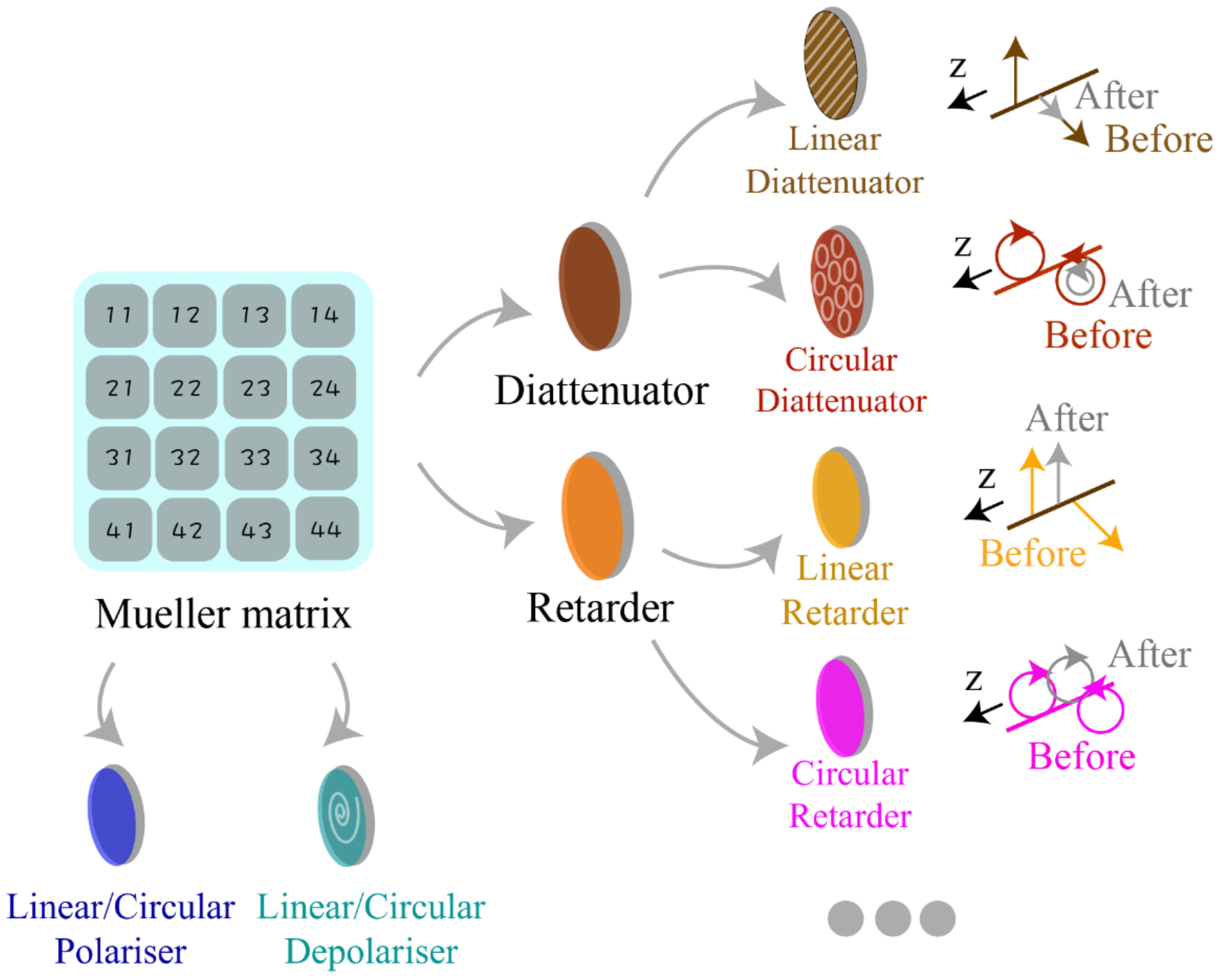

**Supplementary Figure 1. MM and the encoded information.** There are several fundamental polarisation properties that can be extracted via the MM, such as linear/circular diattenuation, linear/circular retardance, and so on. There fundamental optical properties manipulate the light vector along propagation direction z in different format with respect to different eigenbases. They are partially demonstrated, where 'before' and 'after' illustrate the amplitude and/or phase change of the determined eigenvectors.

## Supplementary Note 2: Four vectorial metrics of Mueller matrix

As we have explained in previous note, the widely adopted MM decomposition methods require different assumptions[4-6]. However, there indeed exist vectorial representations whose values are related to physical phenomena that are valid only when specific physical conditions occur (such as for certain multi-layered systems)[3,6-9].

We refer to such representations throughout this paper as 'vectorial metrics', as they reveal certain unique vectorial information. In the main article, we summarised a unified presentation of the metric Δ (first equation) to illustrate the useful quantities extracted from symmetrical or asymmetrical properties of the MM. Furthermore, we explained the how the inference network affects the scope of interpretation of Δ, which is related to information about layered structures in the specimen that can be extracted via the metrics. For instance, if we found the metric values were all zero, we would not be able to determine whether or not such a system is single-layered. As general examples of this scope, we find that certain single-layered structures can infer a zero-valued Δ, whereas the presence of certain multi-layered structures can be inferred through a non-zero value of the Δ. Note that for simplicity in this paper, we confine the scope to single layered systems that are described by a linear, rather than elliptical, eigen basis[10]. For similar reasons, we confine our considerations of depolarisation to homogeneous depolarisation as a single layer[10]. This covers the majority of realistic scenarios and does not preclude extension of the approach to more general cases in later research.

In this note, we detail the four metrics that are used in main article. Supplementary Figures 2a, 2b, 2c and 2d give an overview of the four metrics. Those metrics reveal different physical characteristics of the original complex system using asymmetric properties of the elements of the MM. Such information only exists when the system is complex, in the sense that it consists of multiple layers[3,7-9], as represented by Eq. (1-1). Note that theoretical studies of related parameters have been investigated before with amplification[3,7-9], but they have not yet been summarised into a unified structure, or been put to broad practical use. We summarise metric 1, metric 2, metric 3 and metric 4 in Eq. (2-1) to Eq. (2-4), to act as different quantitative criteria for characterizing vectorial properties in various applications.

Metric 1 is defined as

$$\mathrm{M}_1 = m_{23} - m_{32} \,, \tag{2-1}$$

which relates to the asymmetry between diagonally opposed elements $m_{23}$ and $m_{32}$. If this asymmetry exists, it shows that it is a multi-layered complex system that includes circular retardance[7] (see Supplementary Figure 2a). For the GRIN lens case used in the main text, a non-zero $\mathrm{M}_1$ gradient reflects the existence of circular birefringence gradient in a linear retarder assembly when it is obliquely illuminated. The presence of circular birefringence gradient also validates the existence of spin-Hall effect of the light in this system.

Metric 2 is defined as

$$\mathrm{M}_2 = \sqrt{(m_{12}^2 + m_{13}^2)} - \sqrt{(m_{21}^2 + m_{31}^2)}, \tag{2-2}$$

which represents a different MM asymmetry, relating to polarisance and diattenuation. A non-zero value of $\mathrm{M}_2$ indicates the presence of a multi-layered complex system that includes polarisance and diattenuation. If $\mathrm{M}_2>0$, the diattenuation is stronger than polarisance in the system, and if $\mathrm{M}_2<0$, the polarisance is stronger than diattenuation[3,9]. Assuming we have ideal components, those properties determined by the sign of metric 2 can be equivalently treated as a depolariser followed by a diattenuator (or in reversed order)[3,9] (see Supplementary Note 6). Typically, when either $\sqrt{(m_{21}^2 + m_{31}^2)}$ or $\sqrt{(m_{12}^2 + m_{13}^2)}$ equals to zero, such metric means the existence of polarisance but no diattanuation or vice versa[3,9] (see details in Supplementary Figure 2b);

Metric 3 is defined as

$$\mathrm{M}_3 = \sqrt{(m_{42}^2 + m_{43}^2)} - \sqrt{(m_{24}^2 + m_{34}^2)}, \tag{2-3}$$

which represents a MM asymmetry relating to multiple retarders[7,8]. A non-zero value of $\mathrm{M}_3$ supports the presence of a multi-layered complex system that includes multi-layered retardance structure. Note in the demonstration for metric 3 throughout this paper, our analysis was based on the prior knowledge that scattering and intrinsic stress both exist in fused silica based waveguide systems. Hence, for this fabrication process specifically, this metric can act as an optimiser to balance two types of the retardance with respect to the changing writing parameters (see Supplementary Note 5 and 9).

Metric 4 is defined as

$$\mathrm{M}_4 = m_{14} - m_{41}, \tag{2-4}$$

which relates to the asymmetry between diagonally opposed elements $m_{14}$ and $m_{41}$. A non-zero value of $\mathrm{M}_4$ represents the difference between linear retardance and linear diattenuation (details see Ref[7]), which indicates the presence of a multi-layered complex system. Specifically, if $m_{41} = 0$ but $m_{14} \neq 0$, it shows a double-layered system in which the first layer is a retarder and the second layer is a diattenuator; while for $m_{41} \neq 0$ but $m_{14} = 0$, the order of layers is inverted[7]. Note such conclusions make sense under conditions when the depolarisation is negligible, such as in thin biomedical samples[10-14].

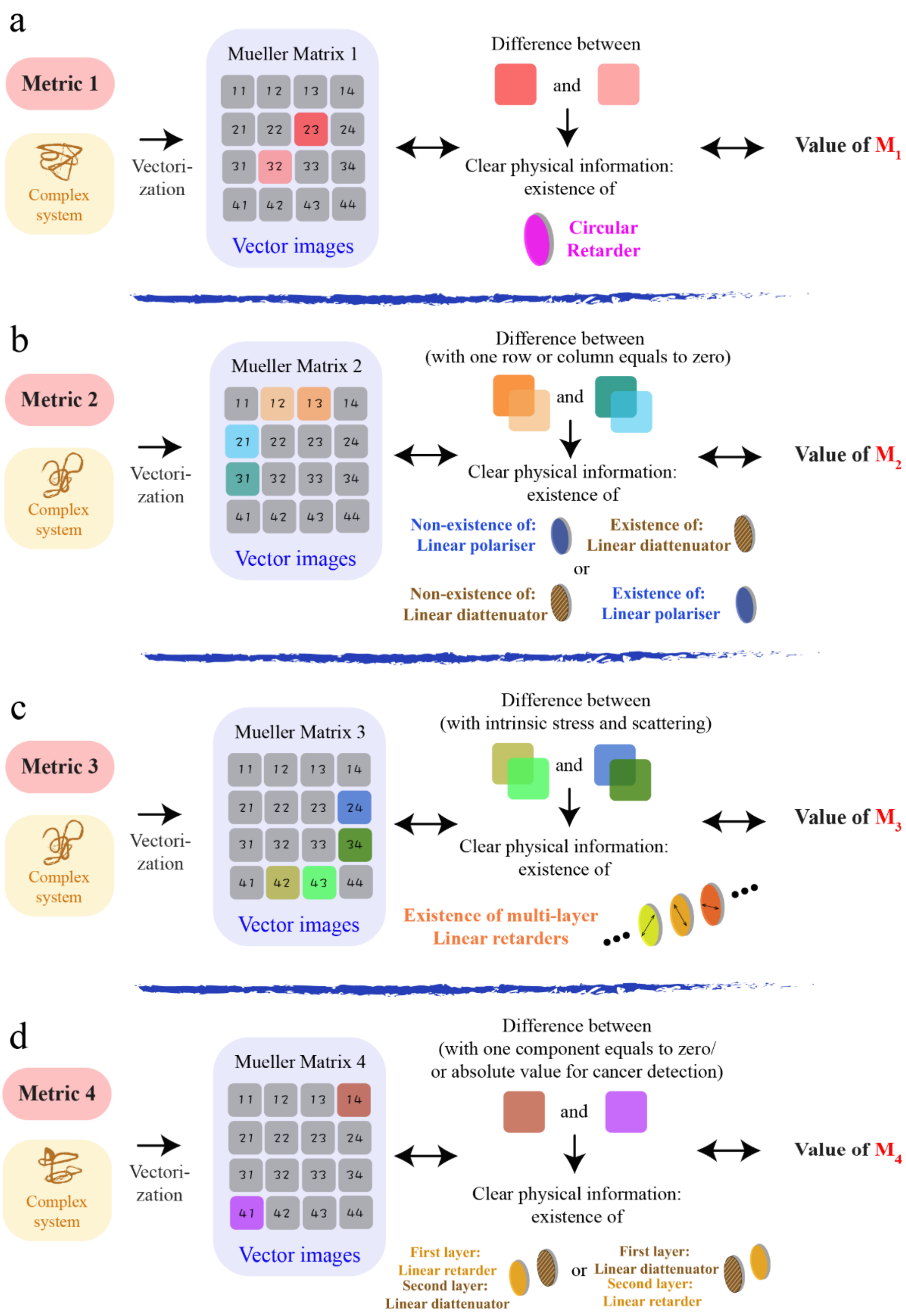

**Supplementary Figure 2. Four vectorial metrics that we used in the main article.** (a) Metric 1 ($M_1$) focuses on the difference between element $m_{23}$ and $m_{32}$, representing the existence of circular retardance. (b) Metric 2 ($M_2$) focuses on the difference between $\sqrt{(m_{21}^2+m_{31}^2)}$ and $\sqrt{(m_{12}^2+m_{13}^2)}$, representing the relationship between polarisance and diattenuation. (c) Metric 3 ($M_3$) focuses on the difference between $\sqrt{(m_{42}^2+m_{43}^2)}$ and $\sqrt{(m_{24}^2+m_{34}^2)}$ representing the condition of multi-layered retardance. (d) Metric 4 ($M_4$) focuses on the difference between element $m_{14}$ and $m_{41}$, representing the sequence of linear retardance and linear diattenuation.

## Supplementary Note 3: Stokes-Mueller measurement and decomposed parameters of a Mueller matrix.

As the MM, its decomposed parameters and Stokes vector fields play important roles in this work, we give a brief introduction of the related measurement methods and the methods of decomposition. Supplementary Figure 3 shows a MM polarimeter using a dual-rotating wave plate method[15]. The polarisers (P1, P2; Thorlabs, LPVISC100-MP2) are fixed and similarly oriented. Two quarter waveplates (QWP1, QWP2; Thorlabs, AQWP10M-580) rotate with fixed rotational speeds, such that $\phi_1 = 5\phi_2$. The main measurement principle is shown in Eq. (3-1) and Eq. (3-2), where $q$ represents the $q^{th}$ measurement. $M_{Sample}$ is the MM of the sample, $M_{P1}$, $M_{P2}$, $M_{QWP1}$ and $M_{QWP2}$ are MMs of P1, P2, QWP1 and QWP2, respectively. $M_{System}$ is the equivalent overall MM of the system. $S_{in}$ and $S_{out}$ are incident and output Stokes vectors. Since the intensity is equivalent to the first element $S_0$ of the Stokes vector, we make $I^q = (S_{out}^q)_0$, which represents the corresponding intensity of the $q^{th}$ measurement. From Eq. (3-2) we can obtain the Fourier series ($a_n$ and $b_n$ are Fourier coefficients, $\phi_1^q$ is the angle of QWP1 at the $q^{th}$ measurement) and calculate the MM, detailed in Ref[15].

$$S_{out}^q = M_{System}S_{in} = M_{P2}M_{QWP2}M_{QWP2}^qM_{Sample}M_{QWP1}^qM_{P1}S_{in}. \tag{3-1}$$

$$I^q = (S_{out}^q)_0 = a_0 + \sum_{n=1}^{12}\left[a_n\cos(2n\phi_1^q) + b_n\sin(2n\phi_1^q)\right]. \tag{3-2}$$

The Stokes polarimeter can be used to calculate any SOP via the sequence of recorded intensities[16]. As we can find in Supplementary Figure 3, the detection arm – also referred to as the polarisation state analyser (PSA) – of the MM polarimeter is a complete Stokes polarimeter. This allows measurement of the Stokes vectors of the light field by rotating the QWP2 to four different angles, following a process described in Ref[16]. The principal equations for calculation of field are shown in Eq. (3-3), in which $S_{in}$ is the Stokes vector of the incident light field, $M_P$ and $M_{QWP}^n$ are MMs of the corresponding polariser and waveplate, respectively. $S_{out}^n$ is the corresponding output Stokes vector for the $n^{th}$ fast axis orientation of QWP2. $M_{QWP2}^n$ is the MM of QWP2 for the $n^{th}$ fast axis orientation. $A$ is a $n\times4$ matrix known as the instrument matrix[17-20], which is derived from $M_P \cdot M_{QWP2}^n$. $I$ is the intensity information recorded by the camera. More practical derivations can be found in Ref[17-20].

$$S_{out}^{n} = M_P M_{QWP2}^{n} S_{in},\ (n = 1,2,3,4 \ldots)$$

$$I = A \cdot S_{in}, \tag{3-3}$$

$$S_{in} = A^{-1} \cdot I.$$

As we mentioned, a MM contains 16 elements, but the relationship between physical quantities and these individual elements is often ambiguous. Previous research has concerned the extraction of specific polarisation parameters from the MM to characterize the optical properties of the object. We introduce two widely used MM decomposition methods in this Note, and use different sub-set of them throughout this work. Even they are limited via different assumptions[4,6], they can still be candidates for assessment of the performance of our metrics. For instance, in the section on biomedical analysis, we use retardance to compare and validate the performance of $\mathrm{M}_4$ (see Supplementary Note 7).

One prevalent method is the MM polar-decomposition (MMPD) proposed by *Lu* and *Chipman*[4], which has been widely used and validated for characterization of biomedical and material samples. In the main article, we specifically use the retardance parameter ($R$), which is suitable for analyzing complex turbid biomedical tissue; this has been applied previously in quantitative biological diagnosis[11-14]. The principle of MMPD is represented by Eq. (3-4), where $\mathrm{M}_\Delta$, $\mathrm{M}_\mathrm{R}$ and $\mathrm{M}_\mathrm{D}$ are the 4×4 sub-matrices of depolarisation, retardance, and diattenuation, respectively.

$$M_{Sample} = M_\Delta M_R M_D. \tag{3-4}$$

The retardance $R$ is reconstructed from the trace of $M_R$; while the orientation of optical axis of linear retardance $\theta$ (with respect to the horizontal axis) ranging from $-\frac{\pi}{2}$ to $\frac{\pi}{2}$ radians are calculated using Eq. (3-5). The diattenuation value $D$ can be readily obtained from the second to the fourth elements in the first row of a MM, while depolarisation properties are included in the bottom right 3×3 matrix $m_\Delta$ of the matrix $M_\Delta$, in which $\lambda_1$, $\lambda_2$ and $\lambda_3$ are the eigenvalues of $m_\Delta$, and $P$ is a matrix composed of the eigenvectors of $m_\Delta$.

$$R = \cos^{-1}\left[\frac{\mathrm{tr}(M_R)}{2} - 1\right],$$

$$\theta = \frac{1}{2}\tan^{-1}\left[\frac{M_{R23} - M_{R32}}{M_{R31} - M_{R13}}\right],$$

(3-5)

$$D = \sqrt{m_{12}^2 + m_{13}^2 + m_{14}^2},$$

$$m_\Delta = P\begin{pmatrix} \lambda_1 & 0 & 0 \\ 0 & \lambda_2 & 0 \\ 0 & 0 & \lambda_3 \end{pmatrix}P^{-1}.$$

*Arteaga et al.* derived MM anisotropy coefficients (MMAC) to describe the degree of different kinds of anisotropy that might be present[6]. Intuitively, the parameters *α*, *β*, and *γ* from MMAC can be regarded as ratios of horizontal linear anisotropy, 45° linear anisotropy and circular anisotropy respectively with respect to the global anisotropy of the MM[6]. Among them, the circular parameter *γ* is used to directly reflect the existence of circular retardance, which is also used as an example to determine the existence of spin-Hall effect of light (SHEL)[21]. We use *γ* to validate the SHEL in a GRIN lens system as supporting evidence for using $M_l$. The mathematical definition of the parameter *γ* with respect to MM elements is shown in Eq. (3-6). For a specific system, if the relative sign between dichroism and birefringence is known, the sign of the MMAC parameters *γ* can be determined as well. Details can be found in Ref[6].

$$\Sigma = 3m_{11}^2 - (m_{22}^2 + m_{33}^2 + m_{44}^2) + 2\Delta,$$

$$\gamma = \sqrt{\frac{1}{\Sigma}[(m_{14} + m_{41})^2 + (m_{23} - m_{32})^2]},$$ (3-6)

$$\mathrm{sign}(\gamma) = \mathrm{sign}(m_{14} + m_{41} \mp (m_{23} - m_{32})).$$

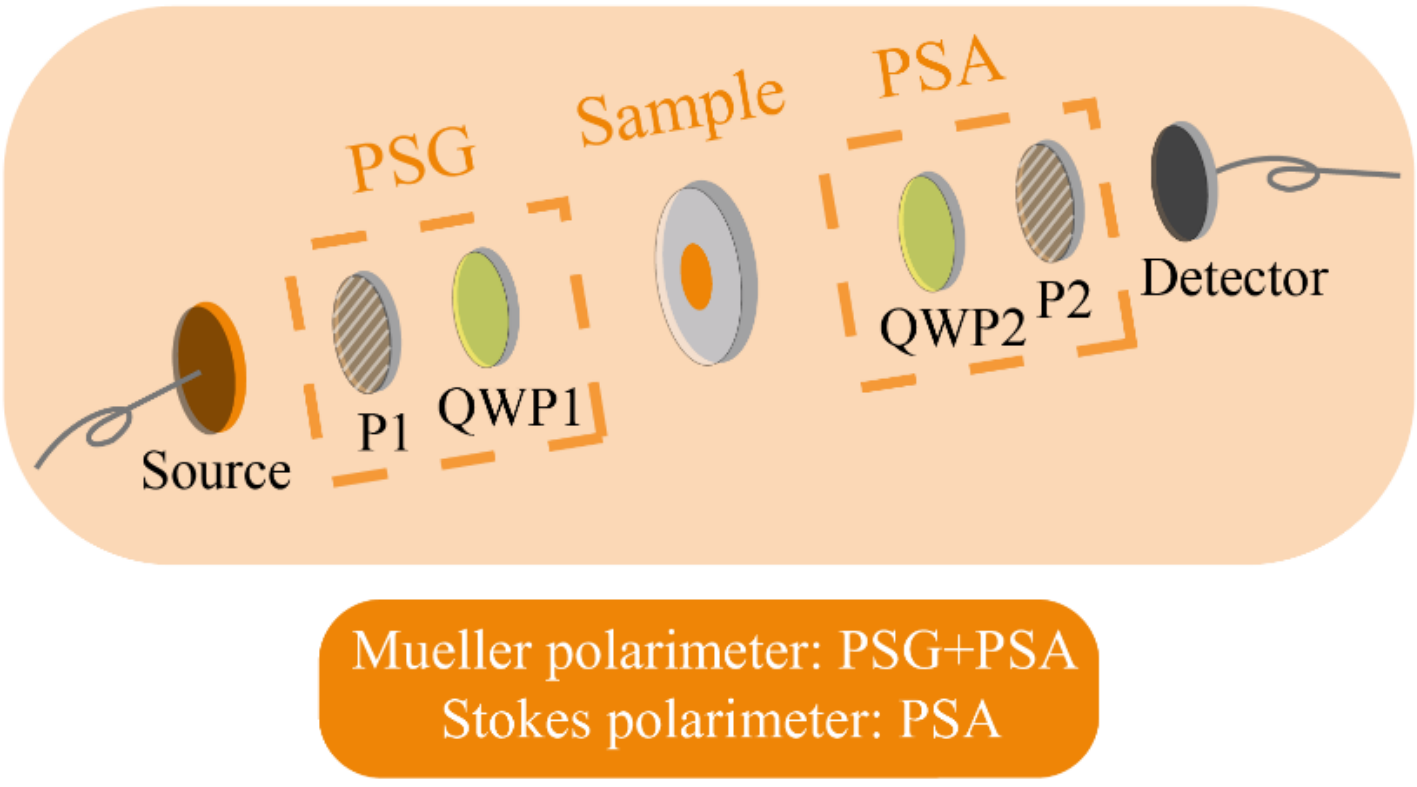

**Supplementary Figure 3. Schematic construction of the MM/Stokes polarimeter.** P1, P2: fixed polariser; QWP1, QWP2: rotating quarter waveplate; a camera is used as the detector. Polarisation state generator (PSG) and polarisation state analyser (PSA) are shown in dotted boxes.

## Supplementary Note 4: Topological charge transfer analysis – from the point of view of angular momentum conservation

It is widely appreciated that in a rotationally symmetric system, the spin orbital interaction (SOI) results in spin-orbital angular momentum conversion (SOC), which normally generates vortex beams[22]. In a system with broken symmetry, the SOI process leads to the spin-Hall effect of light (SHEL), which is directly related to the occurrence of circular birefringence (CB) gradient[21,23]. A GRIN lens has a spatially variant linear birefringence whose fast axis has an azimuthal distribution. It can be treated as a spatially variant waveplate array[18,19,24]. In this work, we break the symmetry of the polarisation aberration of a GRIN lens system by applying collimated, oblique illumination, which experiences a different birefringence to an on-axis beam. Within the numerical aperture (NA) of the GRIN lens, the level of SHEL is determined by the incident angle.

There are various methods to validate the SHEL, such as standard quantum weak measurement methods, Stokes vector analysis, or MM analysis[21,23,25]. The previous two are not suitable for the validation of weak SHEL with complex polarisation modulation, which can be induced by complex Pancharatnam-Berry (PB) phase gradient across the beam transverse plane. Hence, the MM and its decomposition parameters provide unique advantages[21]. They have been used previously for mapping the complex topological structures of the fast-axis orientation of various crystals and validating the appearance of SHEL[21,24].

Here, we demonstrate the SHEL through illuminating a long GRIN lens (Femto Technology Co. Ltd., NA=0.25, 184mm) via an oblique incident angle. Supplementary Note Figure 4 shows the experimental sketch; the incident angle $\Phi$ is between 5° and 10°, with a 1° increment. We measure the full polarisation properties of the GRIN lens through obtaining different MMs for different illumination angles. We then calculate the MMPD and MMAC parameters to analyse the intrinsic mechanism of the related SOI processes. This procedure also acts to validate feasibility for the usage of $M_l$ for identification of SHEL. Figure 5 shows the results of different MMs (see Supplementary Note Figure 5 (a)), different retardance values (see Supplementary Note Figure 5 (b)) and spatially variant fast axis distributions (see Supplementary Note Figure 5 (c)), which are derived from the measured MMs. Note when a tilt is applied, the rotational symmetry of the birefringence distribution is broken (manifested in Supplementary Note Figure 5 (b) and (c)).

The SOI process can be understood from the point of view of angular momentum (AM) conversion. In any complex SOI process in crystals, several topological patterns holding different topological charges are used to assist the descriptions of their AM conversion process – such patterns include the lemon, star, spiral, and node[21]. Suppose we have a right hand polarised photon that carries total AM (TAM) J = +1ħ, consisting of spin angular momentum (SAM) σ = +1ħ and OAM l = 0ħ. When it passes a GRIN lens, the interaction in the node topological area in Supplementary Note Figure 5 (c) converts SAM handedness of the photon into σ = -1ħ. The TAM is changed by a SAM induced factor +2ħ, where the SAM completely transfers to an intrinsic OAM (IOAM). This would significantly change the SOI process, since there exist three different topologies rather than a pure node topological pattern. From Supplementary Note Figure 5 (c) (when the incident angle is 9°) we could clearly find two lemons (red circles), a node (green ellipse), and one linear gradient (black rectangle) at the intermediate region. For the node region, the SOC performance is discussed above. For the lemon regions (which are newly formed), they would gain an IOAM of +1ħ. So, considering the whole field, due to the TAM conservation mechanism, the remaining +1ħ is expected to transfer into the medium – which is also comparable with the SOI process in KDP[21]. The linear gradient areas (black rectangle) are not associated with any azimuthal/radial coordinate. For such a beam-field, the photons passing through this region would neither obtain IOAM nor involve any SOC. Hence, from the point of view of AM conservation, such a region could contribute to extrinsic OAM hence validating the existence of SHEL.

The level of SHEL is quantified via γ at different illumination angles. The process is as follows: 1) we chose three fixed concentric circles on the GRIN lens surface (shown in Supplementary Note Figure 6 (a) with dotted rectangles) near the linear gradient region; 2) at every angle, we obtained the data from the absolute value of γ of the three rectangles; 3) we calculated the mean value and standard deviation through statistical methods[26]; 4) we collected data from different angles and formed the Supplementary Note Table 1 and Figure 6 (b), which show that γ increases with incident angle. To the best of our knowledge, this is the first report of SHEL in a GRIN lens system, as revealed via MM analysis. As the GRIN lens still maintains its basic function as an imaging lens, there exists potential that the GRIN lens can combine imaging and SHEL detection.

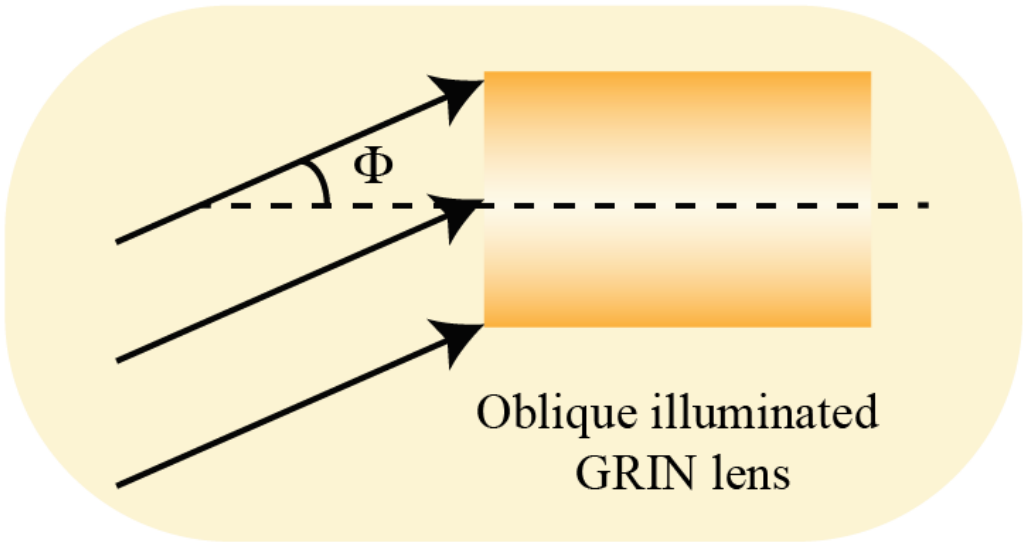

**Supplementary Figure 4. Sketch of GRIN lens oblique illuminated by collimated light beam.** The incident angle $\Phi$ was varied between 5° and 10° as MMs were recorded.

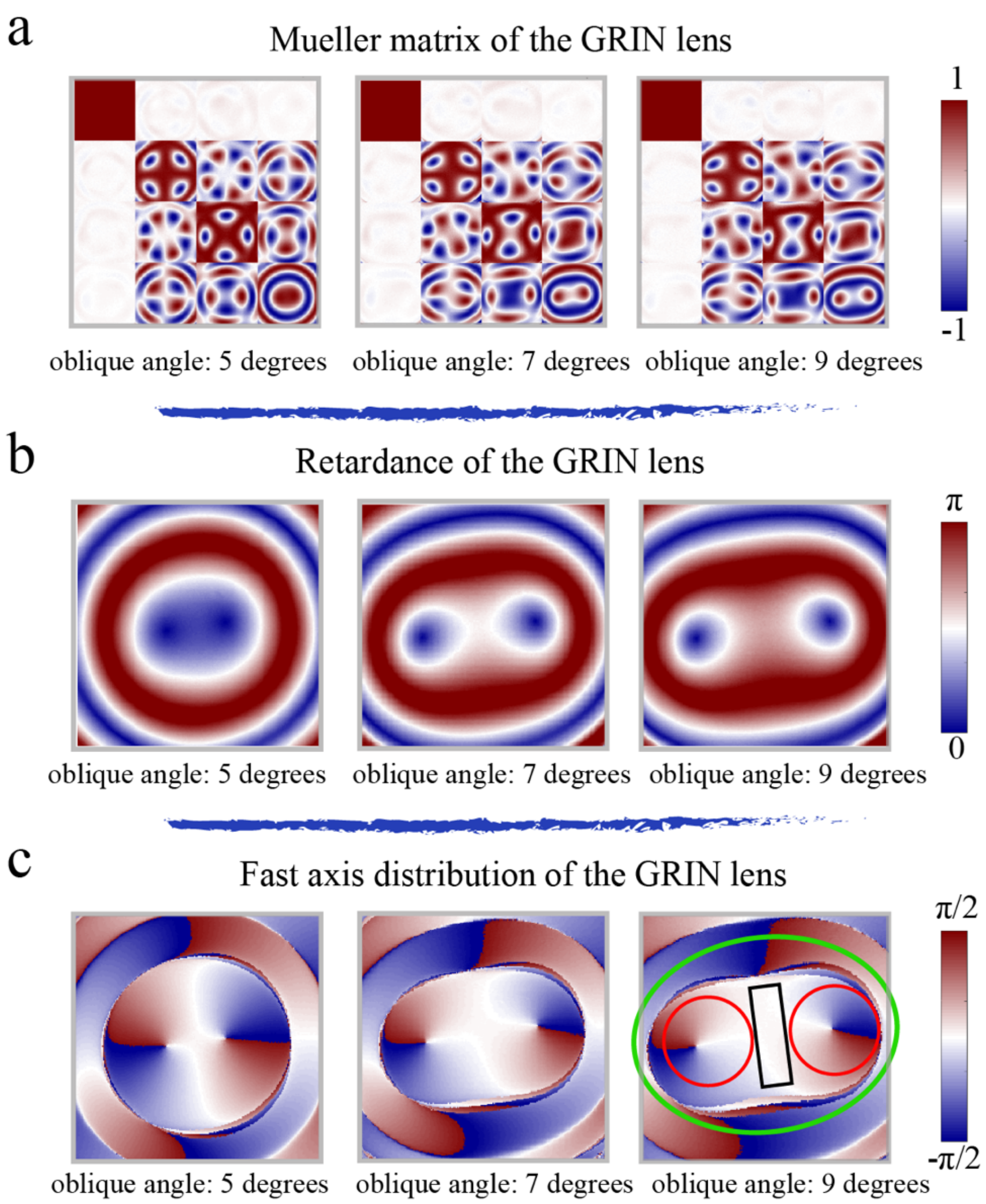

**Supplementary Figure 5. Experimental results of MM and decomposed parameters under different illumination angles.** (a) MMs of GRIN lens with different angle; (b) retardance properties of the GRIN lens; (c) fast axis properties of the GRIN lens. They are demonstrated at incident angle $\Phi$ of 5°, 7° and 9°. The green circle, red circle and black box rectangle represent different topological charge units.

**Supplementary Table 1. Circular anisotropy coefficient (γ) value under different illumination angles**

| | 5° | 6° | 7° | 8° | 9° | 10° |
|---|---|---|---|---|---|---|
| Mean value | 0.083 | 0.128 | 0.133 | 0.166 | 0.209 | 0.234 |
| Standard derivation | 0.021 | 0.026 | 0.018 | 0.024 | 0.025 | 0.019 |

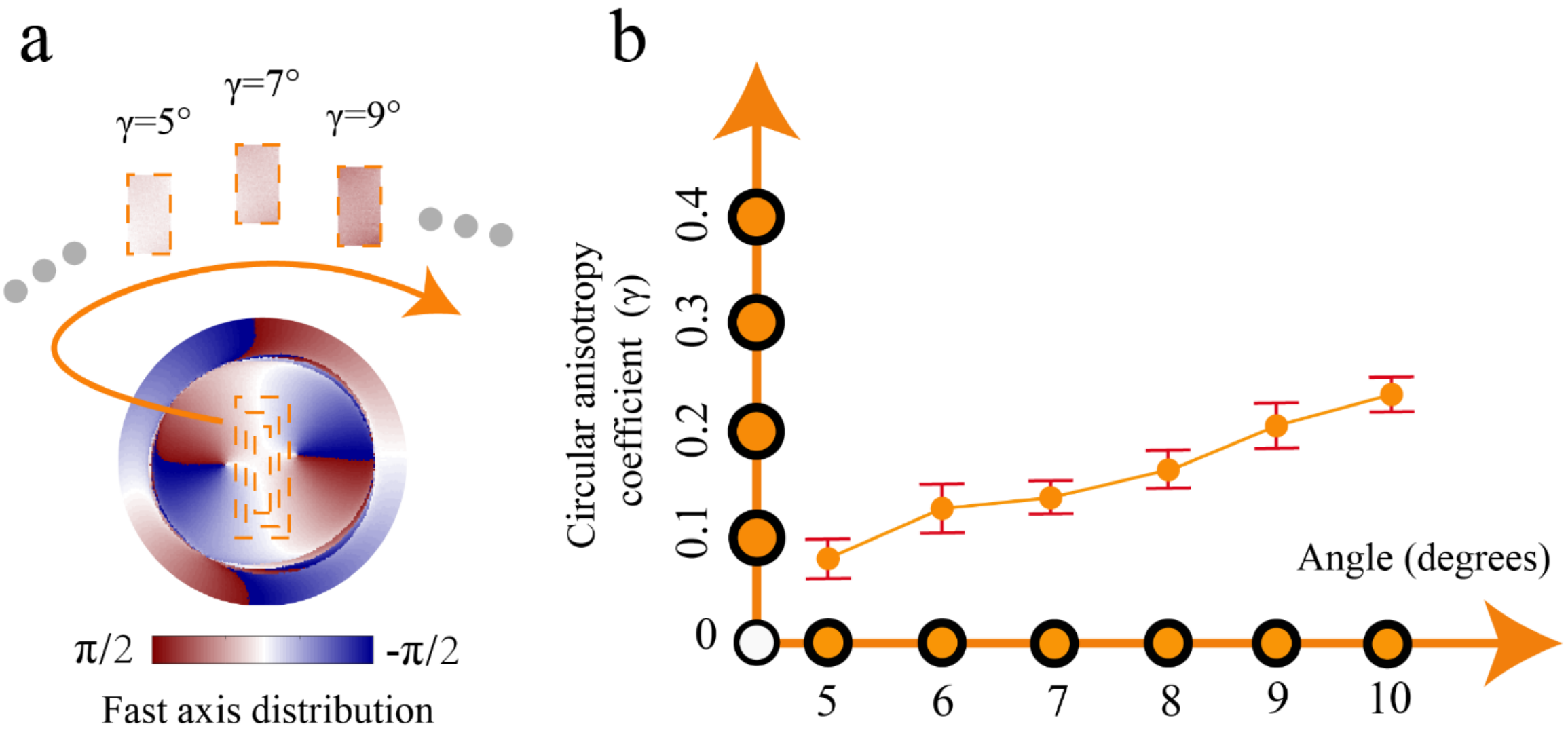

**Supplementary Figure 6. Parameter γ extracted from the measured MMs.** Value of parameter γ extracted under different illumination angles. (a) The three fixed rectangle regions chosen from same physical locations on GRIN lens surface for γ value estimation for each determined illumination angle. (b) The solid yellow circle points and the red error bars represent the mean values and the standard deviation of the γ of three rectangles that were measured at every illumination angle.

## Supplementary Note 5: Direct laser writing of photonic waveguides

Direct laser writing was used to manufacture the photonic waveguides that were studied through metric 2 in the main article. The waveguides were written inside fused silica (Schott Lithosil Q0) with a tightly focused ultrashort pulse laser (Yb:KGW laser, Light Conversion Pharos SP-06-1000-pp; 1 MHz repetition rate; 514 nm wavelength; 170 fs pulse duration). The laser was focused 300 µm below the top surface of the glass with a 0.5 NA objective lens, and its power was controlled by a rotating half waveplate in conjunction with a polarisation beam splitter (PBS). A liquid crystal spatial light modulator was used to compensate system and sample aberrations encountered during processing, as described in Ref[27,28]. The fused silica chip was mounted on a 3-axis precision translation stage (Aerotech ABL10100L/ABL10100L/ANT95-3-V). Waveguides were written by scanning transverse to the optic axis at a speed of 2 mm/s. The writing pulse energy was measured *in situ* at the sample plane, with six values of 35nJ, 42nJ, 50nJ, 58nJ, 67nJ, 75nJ, which were used in the analysis in the main article.

In case 2 in the main article, the value of $M_3$ indicated the existence of a multi-layered linear retarder assembly. It is known that retardance is induced here through scattering and intrinsic stress, so the information we extract is only indicative of average waveguide performance. However, such an approach also presents new possibilities for an optimized fabrication process; e.g., if the value of $M_3$ vanishes through changing the writing parameters, it may prove to be suitable for feedback control, showing an overall retardance performance inside the complex waveguide. In polarisation optics, matrix reciprocity always exists in a multiply layered system, and hence leads to different polarimetric properties observations when flipping the specimen[1-3]. Since the asymmetric MM elements were observed in main article for waveguides under the above writing parameters, we also wrote identical waveguides bi-directionally to test this hypothesis (see details in Supplementary Figure 7a). The linear retardance value of those waveguides was decomposed and is shown in Supplementary Figure 7b and 7c for the six different pulse energies. The results are symptomatic of the quill effect[29,30], where the laser induced structural modification inside an isotropic material is different when reversing the writing direction. Here we additionally confirm the directionality in the fabrication from a polarisation optics standpoint (compared with existing methods[29,30]) using the different small retardance values that are provided via MM images. The same laser writing parameters lead to different linear retardance for opposite writing directions. The statistical estimation approach is shown in Supplementary Note Figure 7e; we chose three fixed circle regions within the aperture (shown in dotted circles, data collected within the circular 2D regions) for statistical approach refer to Ref[26]. Note the waveguides were written in opposite directions but measured in the same orientation; the sample orientation was also flipped in testing for the quill effect.

To the best of our knowledge, it is the first time that the quill effect (and the corresponding trends) in direct laser written waveguides has been investigated from the point of view of the asymmetric information and decomposed polarisation parameters from the MMs. Note here that we used MMPD methods to extract the linear retardance simply for validation and comparison, but bear in mind the complex intrinsic isotropic/anisotropic scattering-induced vectorial manipulation for the beams are still to be explored – as it would affect the intensity loss and polarisation independence as well, in conjunction with the effects from the stress induced retarders. There is intriguing scope for future work.

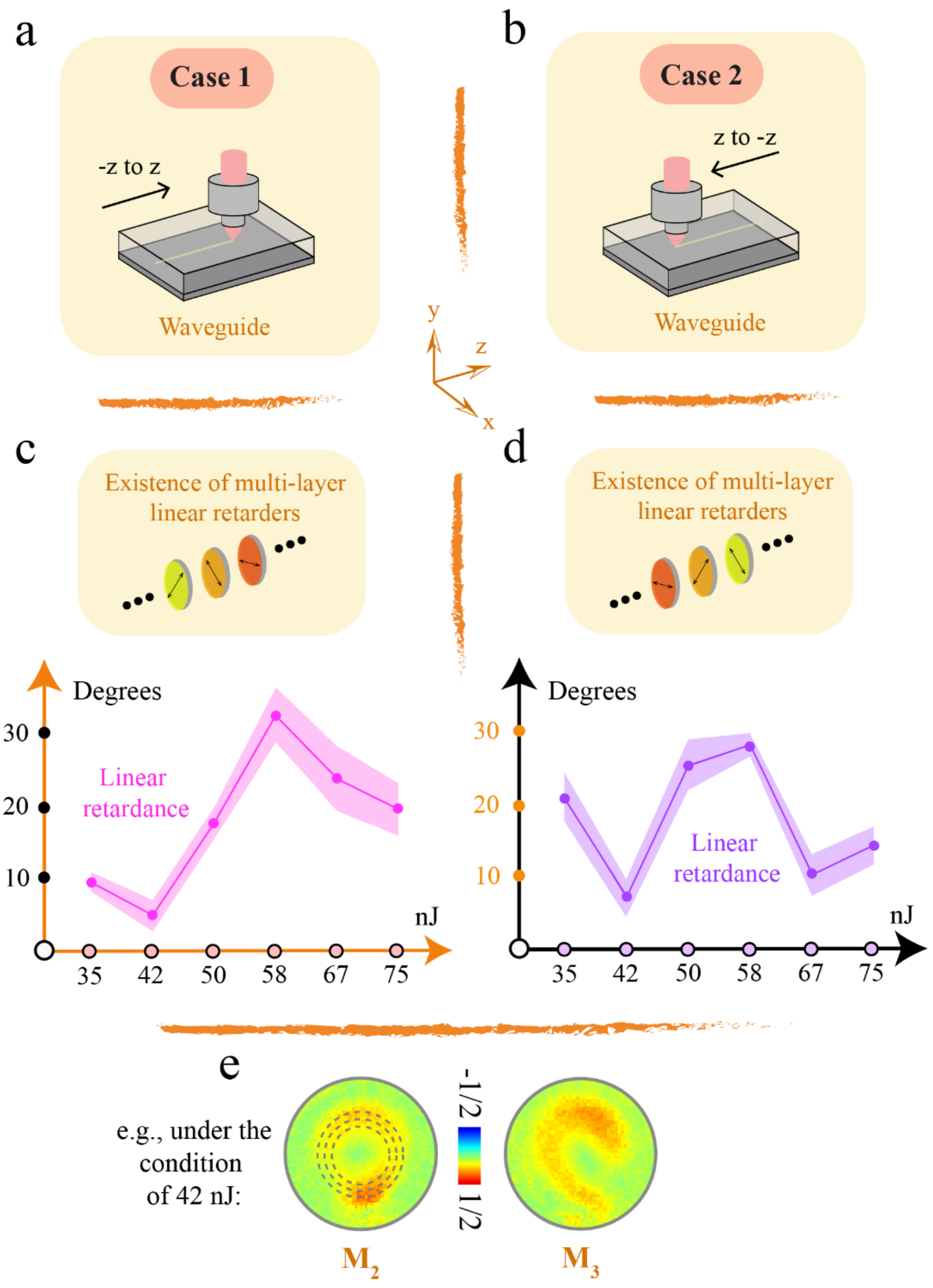

**Supplementary Figure 7. Two direct laser writing cases with different writing directions.** (a) and (b), sketches for the two cases. (c) and (d) the measured linear retardance for the two direction cases with different writing powers. (e) Regions chosen for statistical analysis (data within three fixed dotted circles were used in the plot of the value of various polarisation properties for illustration). Taking retardance as an example, as in (c), the mean value of the dataset is shown by the solid pink circle points, while the transparent pink regions represent the standard deviations.

## Supplementary Note 6: Monte Carlo simulation of intrinsic waveguide optical processes

The waveguides were written in fused silica, which has no inherent polarisation properties. Hence, the polarisation modulations that were observed are due to material changes induced by the direct laser writing process[30-32]. Previous studies have validated that nano-structures[33-39] can be formed in such processes inside fused silica, and that the optical properties change with laser pulse energy (and other parameters), as does the related scattering. Considering that anisotropic scattering and the isotropic scattering both exist in the waveguides, for a better understanding of the experimental results, we carried out Monte Carlo (MC) simulation[10,40,41] with a sphere-cylinder birefringence model (SCBM) to simulate the interactions between polarised photons and the nano-structures of the waveguide. In the SCBM model, the spherical and infinitely long cylindrical scatterers can provide isotropic and anisotropic scattering effects, respectively. Previous studies have also indicated that the anisotropic scatterers can generate linear diattenuation (anisotropic absorption)[10], which can be extracted quantitatively using MM elements. To study the propagation of polarised light in a waveguide, during the MC simulations the scattering coefficients, refractive indices, sizes of both scatterers, the orientation and angular distribution of the cylindrical scatterers, together with the value and fast axis orientation of birefringence for interstitial medium can all be adjusted. Here in this study, the following simulation parameters were used: a two-layered medium (case 1 in Supplementary Figure 8), four-layered medium (case 2), six-layered medium (case 3); case 2 and case 3 have periodic double-layered medium in case 1, which fit the periodic structure that occurs inside the waveguides[35-37]. The thickness of overall layer was 13.6 mm to match the fused silica sample that was used in the experiment. In each double-layered structure: layer 1 consisted of well-ordered cylindrical scatterers with scattering coefficient and size 1 $cm^{-1}$ and 0.01 μm, respectively; the cylinders were distributed along the X axis direction with 5° fluctuations. Layer 2 consisted of spheres and birefringent interstitial medium, the scattering coefficient and size were 15 $cm^{-1}$ and 0.3 μm, respectively; the value and fast axis orientation of birefringence were set to be $\Delta n=0.002$ and 0°, respectively.

Supplementary Figure 8 shows the MC simulated results, which contain polarisance as well as diattenuation and retardance. Such phenomena can also be explained via a layered structure model: scattering-induced depolarisation, followed with a scattering-induced diattenuator, see Supplementary Figure 8 (a). From the simulated results shown in Supplementary Figure 8 (b), we can observe that with the direction of cylinders changes from 0° to 180°. For each case, the value fluctuations of $m_{12}$ and $m_{13}$ are more prominent than those of $m_{21}$ and $m_{31}$, indicating that diattenuation is more dominant than polarisance for these cases. Meanwhile, the values of $m_{42}$ and $m_{43}$ are also larger than those of $m_{24}$ and $m_{34}$, showing the existence of layered linear retardance. Supplementary Figure 8 (b) demonstrates different numbers of the layered structures consisting of the units shown in Supplementary Figure 8 (a), which mimics the periodic structure that occurs inside the waveguides, as mentioned before.

The MC simulation built a link between the measured data of the waveguides as well as the possible polarisation properties models. Given the amplitudes and trends of the elements in $M_2$ and $M_3$ using MC simulation, we also validated the usefulness of our vectorial metrics.

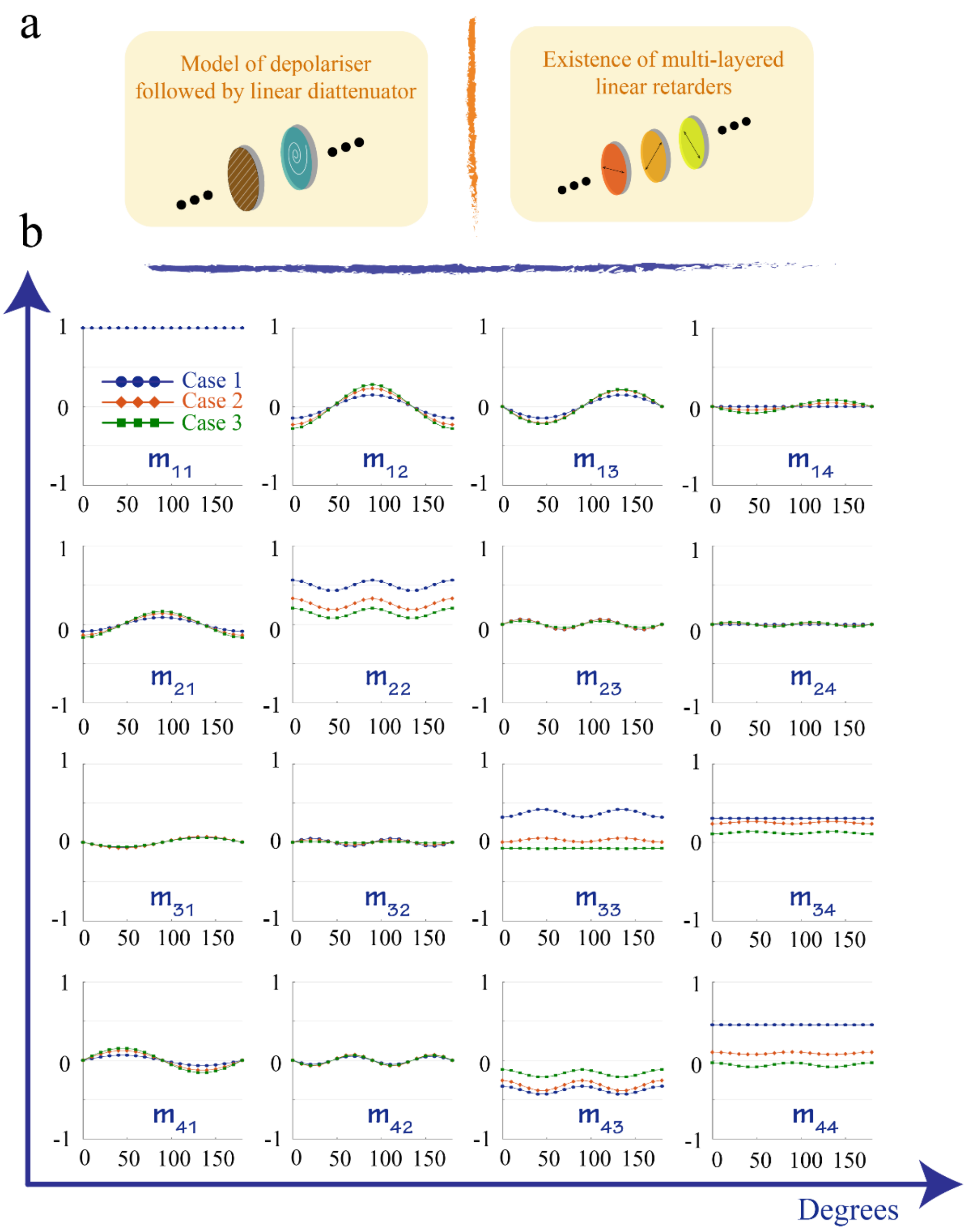

**Supplementary Figure 8. Monte Carlo simulation for the optical properties of the waveguides.** (a) The model of a linear diattenuator and a depolariser (isotropic and anisotropic scattering induced), as well as a multi-layered retardance geometry. (b) Azimuthal dependence curves of the MM elements for SCBM. All the MM elements are normalized by $m_{11}$. Cases 1 to 3 consist of different layered structures, as explained in the text.

## Supplementary Note 7: Data acquisition, data processing, and statistical analysis for non-small cell lung carcinoma

Lung cancer is one of the most commonly diagnosed cancers, and accounts for more than 20% of all cancer deaths worldwide[42]. Clinically, non-small cell lung carcinoma is the primary form of lung cancer constituting about 85-90% of all lung cancer cases[43,44]. Fibrous structures of the extracellular matrix play an important role in the development of non-small cell lung carcinoma tissues since they provide strength and cushioning[45,46]. Recent studies showed that the proportions and distribution of such fibrous structures are different between normal and cancerous lung tissues[45,46]. These differences can be identified with hematoxylin and eosin (H and E) staining. However, for quantitative and accurate evaluation of detailed structural changes of fibrous structures in non-small cell lung carcinoma tissues, MM microscopic imaging was used in this study. Here we selected five 12-μm-thick non-small cell lung carcinoma tissues slices for demonstration. For pathological comparison, the corresponding 4-μm-thick H and E stained slices were also prepared. The sample selection and preparation were performed by experienced pathologists. The age range of the patients was from thirty to fifty-five years. This work was approved by the Ethics Committee of the Shenzhen Second People's Hospital.

Measurements were taken using a conventional MM microscope and the retardance was derived via MMPD, a method that has been used for cancer differentiation in various works[11-13]. We selected 10 points per region; 2 regions were chosen from a sample, one containing normal and one containing cancerous tissue. The statistical analysis followed the approach of Ref[26] (Supplementary Figure 9a). The field of view (FOV) of the MM microscope was 0.77 $mm^2$ and its calibrated precision was lower than 0.3%. We calculated the mean value and standard deviation of the retardance across the same areas to set as ground truth for comparison with the absolute value of metric 4. Example MMs, as well as corresponding MMPD parameters, $M_4$ value from healthy or cancerous tissue and quantitative statistic histograms are illustrated in Supplementary Figure 9.

Supplementary Table 2 contains the data measured by two parameters as well as the corresponding P-value, which shows the significant difference between the two classes of samples. Considering the original data distribution and the P-values of either individual samples or the overall combination, it can be found that the metric 4 is able to distinguish efficiently between healthy and cancerous tissue. To make the process more efficient and precise, further detailed error analysis and corresponding optimization will be the subject of further work.

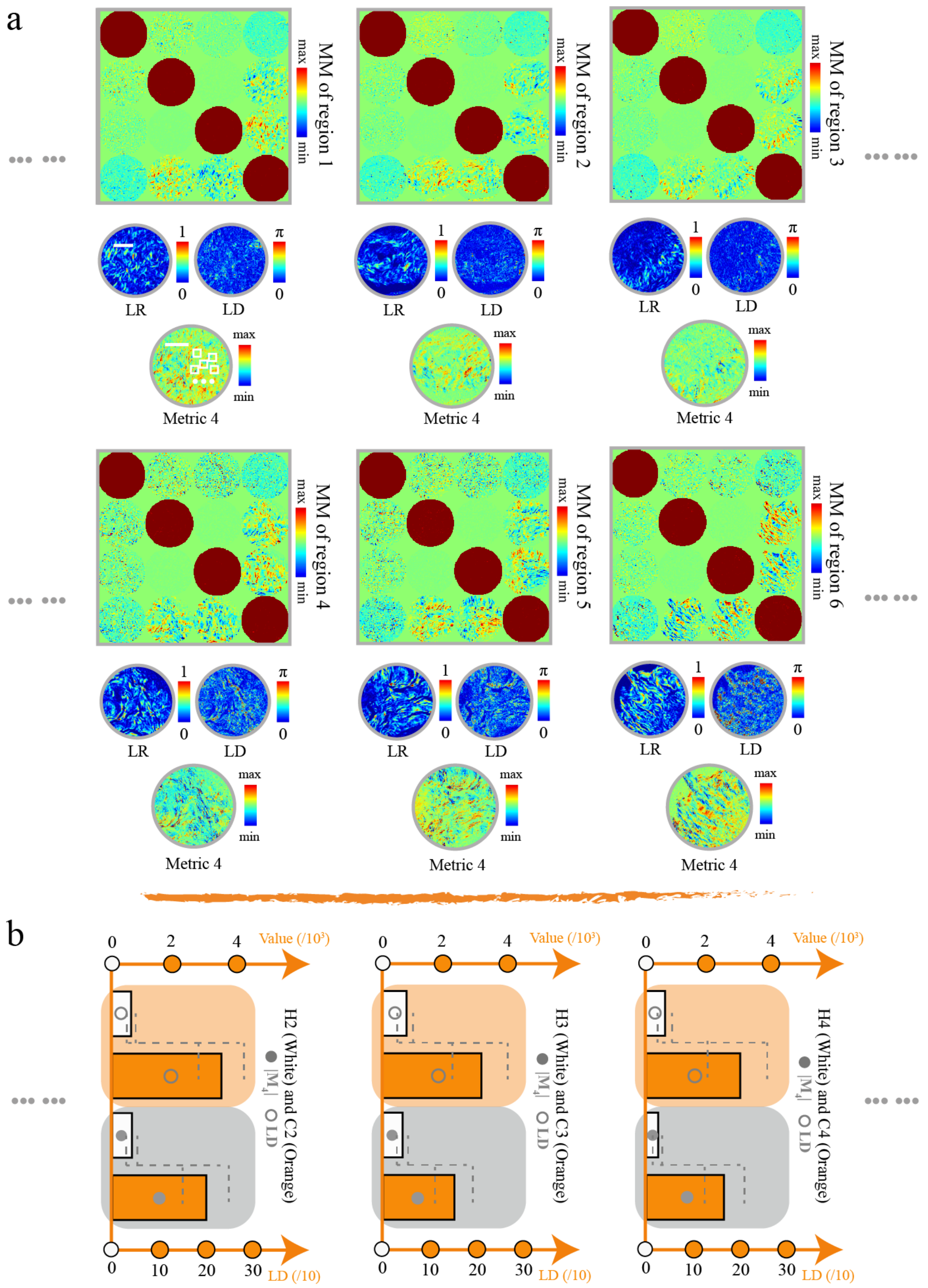

**Supplementary Figure 9. Demonstration of original MMs, decomposed parameters, value of metric 4 and statistical histograms of healthy and cancerous lung tissue.** (a) MMs and related parameters – alongside with the value of linear diattenuation (LD), linear retarder (LR), metric 4; Note the scale used for LD, LR, and $M_4$ has been amplified by a factor of 4, 5 and 20 for better visualisation[26]. (b) Statistical histograms of some samples (note here we use the absolute value of metric 4). Gray dotted lines represent the standard deviation.

**Supplementary Table 2. Value of retardance and metric 4 of the samples used in this work**

**(Retardance: $/10^1$; $M_4$: $/10^3$)**

| | Slide 1 (H1 and C1) | | | | | Slide 2 (H2 and C2) | | | | |
|---|---|---|---|---|---|---|---|---|---|---|
| | *Healthy* | | *Cancerous* | | | *Healthy* | | *Cancerous* | | |
| | *mean value* | *standard deviation* | *mean value* | *standard deviation* | *P-value* | *mean value* | *standard deviation* | *mean value* | *standard deviation* | *P-value* |
| **Retardance** | 3.78 | 1.56 | 16.22 | 4.23 | <0.001 | 4.51 | 1.42 | 19.31 | 5.02 | <0.001 |
| **$\lvert M_4 \rvert$** | 0.39 | 0.21 | 2.84 | 0.77 | <0.001 | 0.32 | 0.09 | 3.75 | 0.86 | <0.001 |
| | **Slide 3 (H3 and C3)** | | | | | **Slide 4 (H4 and C4)** | | | | |
| | *Healthy* | | *Cancerous* | | | *Healthy* | | *Cancerous* | | |
| | *mean value* | *standard deviation* | *mean value* | *standard deviation* | *P-value* | *mean value* | *standard deviation* | *mean value* | *standard deviation* | *P-value* |
| **Retardance** | 4.56 | 1.39 | 15.73 | 3.97 | <0.001 | 2.24 | 0.82 | 17.55 | 3.34 | <0.001 |
| **$\lvert M_4 \rvert$** | 0.42 | 0.19 | 3.13 | 0.89 | <0.001 | 0.24 | 0.11 | 2.92 | 0.93 | <0.001 |
| | **...** | | | | | **Combination of all slides** | | | | |
| | *Healthy* | | *Cancerous* | | | *Healthy* | | *Cancerous* | | |
| | *mean value* | *standard deviation* | *mean value* | *standard deviation* | *P-value* | *mean value* | *standard deviation* | *mean value* | *standard deviation* | *P-value* |
| **Retardance** | … | | | | | 3.39 | 1.15 | 17.31 | 4.07 | <0.001 |
| **$\lvert M_4 \rvert$** | … | | | | | 0.29 | 0.13 | 3.29 | 0.82 | <0.001 |

## Supplementary Note 8: Monte Carlo simulation for double-layered system

As shown in the Fig. 4d in the main article, as well as Supplementary Note 2, the zero or non-zero values of $m_{14}$ or $m_{41}$ represent different layered structural information of tissues. Previous studies have demonstrated that such layered structures exhibit different linear diattenuation and retardance properties[7,8], which may be helpful for biomedical measurements. Here, for a better understanding of the experimental results, we carry out MC simulation[40,41] with a cylinder birefringence model (CBM) to simulate the interactions between polarised light and the fibrous structures of tissues. In the CBM model, the infinitely long cylindrical scatterers can provide similar anisotropic scattering effects and linear diattenuation to that generated by the tissue fibers. During the MC simulations the scattering coefficients, refractive indices, sizes, orientation, angular distribution of the cylindrical scatterers, and the value and fast axis orientation of birefringence for interstitial medium can be adjusted. Here in this study, the simulation parameters were set as follows. For the two-layered medium, the thickness of both layers was 0.006 mm; for the layer consisting of scattering cylinders, the scattering coefficient and size were 200 $cm^{-1}$ and 0.05 μm, respectively; the cylinders were distributed along the X axis direction with 5° fluctuations; for the layer of birefringent medium, the value and fast axis orientation were set to be $\Delta n$=0.002 and 22.5°, respectively.

Supplementary Figure 10(a) gives three cases that we simulated in this section, which represent different morphologic geometries in such human lung samples. From the simulated results shown in Supplementary Figure 10 (b) we can observe that the $m_{14}$ and $m_{41}$ elements show asymmetric properties, especially for cases 2 and 3, in which the value of $m_{14}$ (or $m_{41}$) has a more prominent fluctuation than that of the $m_{41}$ (or $m_{14}$) with the direction change of cylinders. For the medium in case 1 combining cylinders and birefringence in the same layer, this asymmetric property occurs in both elements. Meanwhile, other MM elements of such a two-layered medium are symmetrical.

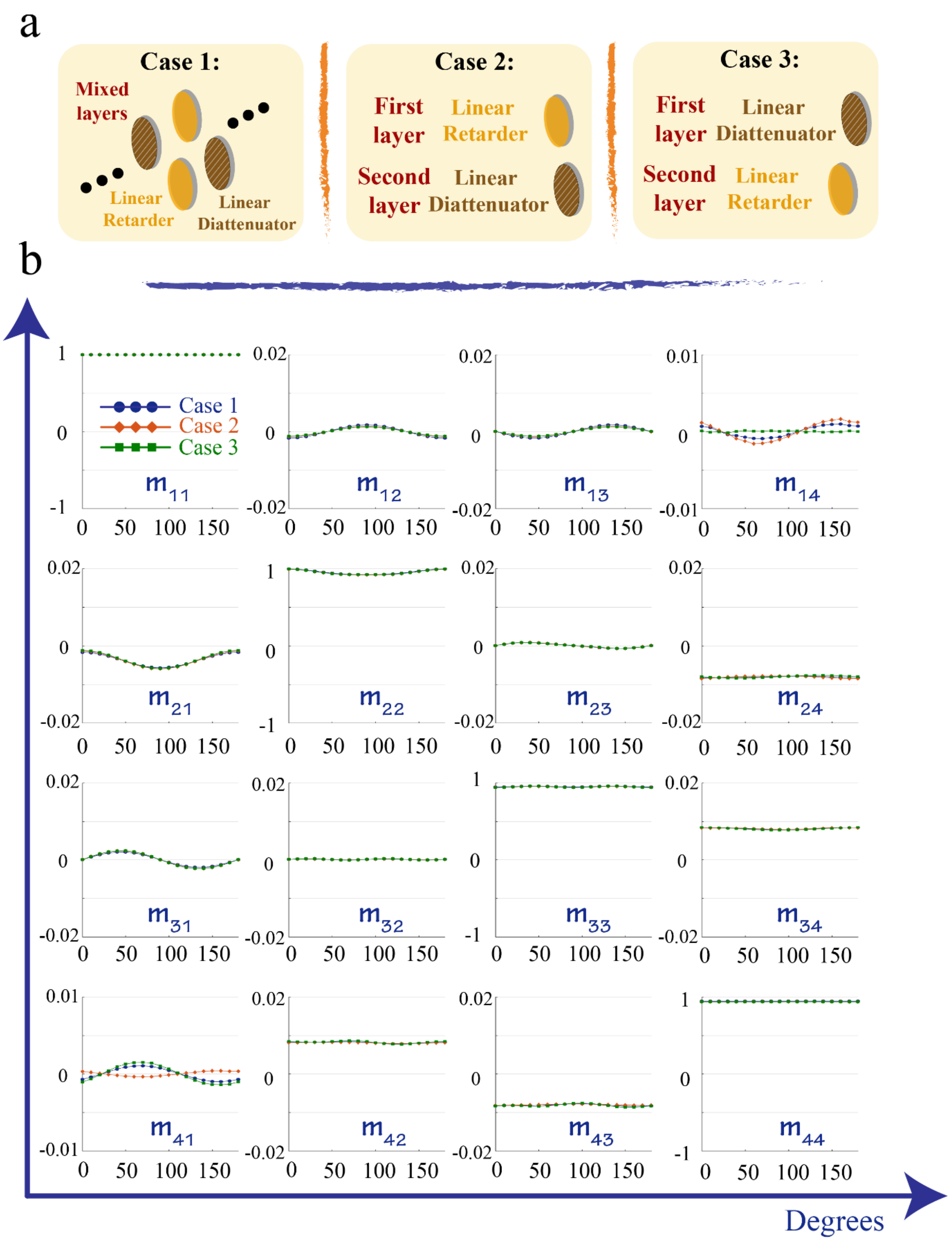

**Supplementary Figure 10: Monte Carlo simulation for the optical properties of a biomedical sample.** (a) The model of case 1 (mixed-layered geometry), case 2 (first layer: linear retarder, second layer: linear diattenuator), as well as case 3 (first layer: linear diattenuator, second layer: linear retarder). (b) Azimuthal dependence of the MM elements for CBM. All the MM elements are normalized by $m_{11}$.

## Supplementary Note 9: More vectorial metrics derived from the Mueller matrix

The four previously introduced metrics (1, 2, 3 and 4) are based on analysing the asymmetric or symmetric properties of the MM images; these could be referred to as "difference metrics" that have the form (X-Y), where X and Y are each MM elements (or combination of elements). The new proposed metrics (metric 5 in the discussion) could be defined as a "ratio metric" in the form (X/Y) where X is the MM element (or combination of elements) that can be expected to be near-zero; Y is a reference value derived from the element $m_{11}$.

Through exploration of various cases, we can categorize the different metrics according to the unified map in the figure (Supplementary Figure 11a). We mainly present three general types of metric in the note: element-wise, column/row-wise and block-wise. Specifically, for block-wise, as the MM is a 4 x 4 matrix, the sub-block would be an n x m sub-matrix ($1<n, m<4$). Where one metric belongs to above categories. To form a new metric, we may choose any combination of elements; we do not need to be restricted to symmetrically opposed elements (see Supplementary Figure 11a).

Some new vectorial metrics are defined in Supplementary Figure 12, which additionally reveals six potential metrics (metric 6 to 11) beyond the ones that have been explained in the main article. Supplementary Figure 13 shows two pairs of GRIN lens cascades with theoretical and experimental data as validation examples. These cases have been illustrated using a spatially variant half-wave plate array (SVHWP) based GRIN lens cascade[24]; their MMs and certain vectorial properties are shown in Supplementary Figure 14. Here we provide just a brief illustration of the possibilities for other metrics that can be further developed within the scope of Supplementary Figure 11a.

We then give a brief summary in Supplementary Figure 11b about the functionalities of the vectorial metrics that appeared in this paper. They can act as indicators to 1) reveal information of complex optical systems (e.g., metric 1-5); 2) optimize operation such as a feedback metric for control purposes (e.g., metric 2, 3 and 5); 3) suppress values in the MM (e.g., metric 6 to 11, maintain various elements whose absolute values have been suppressed). The third functionality, which is revealed here, may provide an intriguing possibility to edit the presence of a particular physical property, or do vectorial coding/decoding of the MM itself via its sixteen elements. The overall picture shows the opportunity for further investigation of such vectorial metrics.

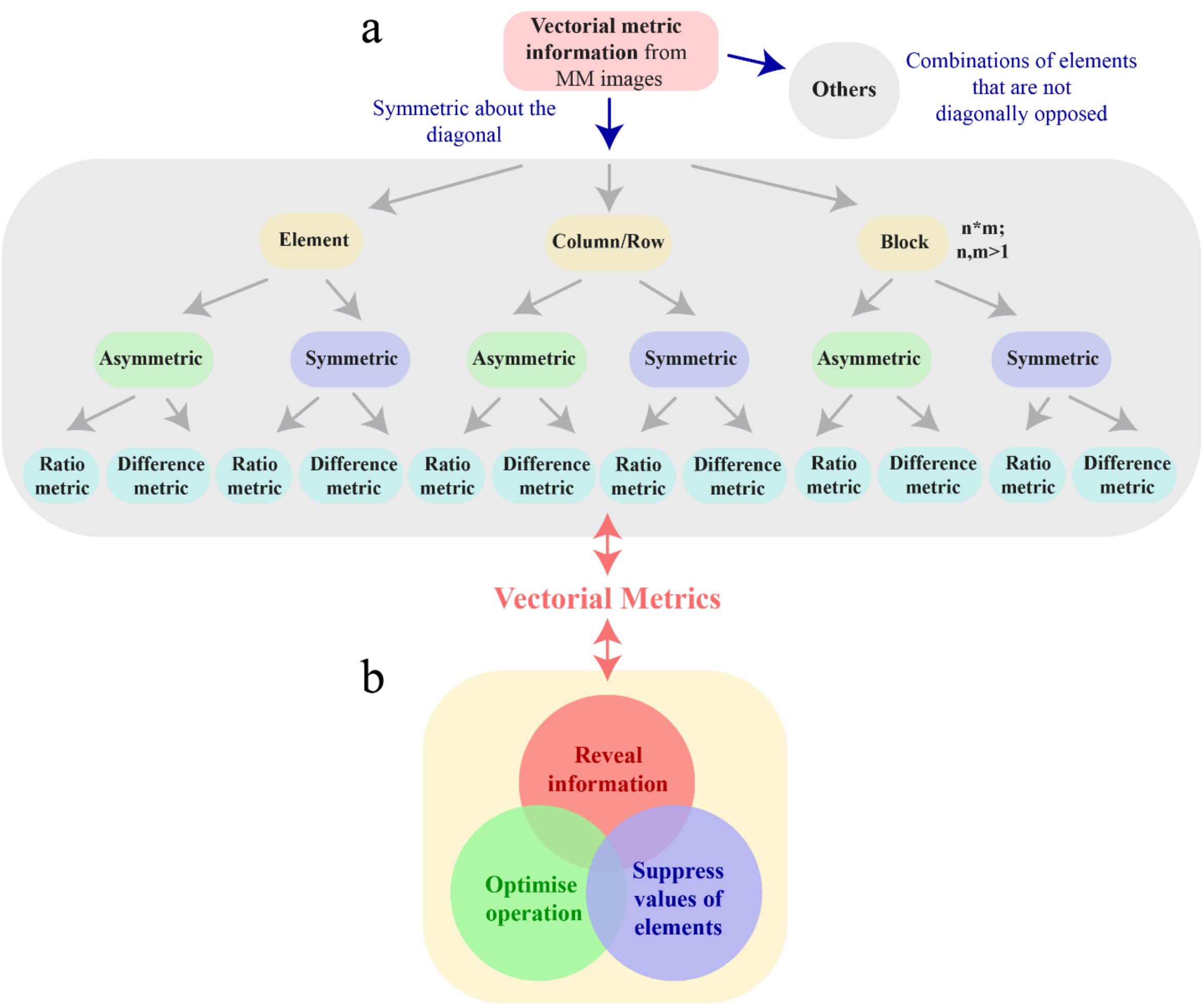

**Supplementary Figure 11. Vectorial metric information from MM images.** (a) The picture gives an overview of different categories of the information extraction from the MM vector images. (b) Three characteristics of the vectorial metrics.

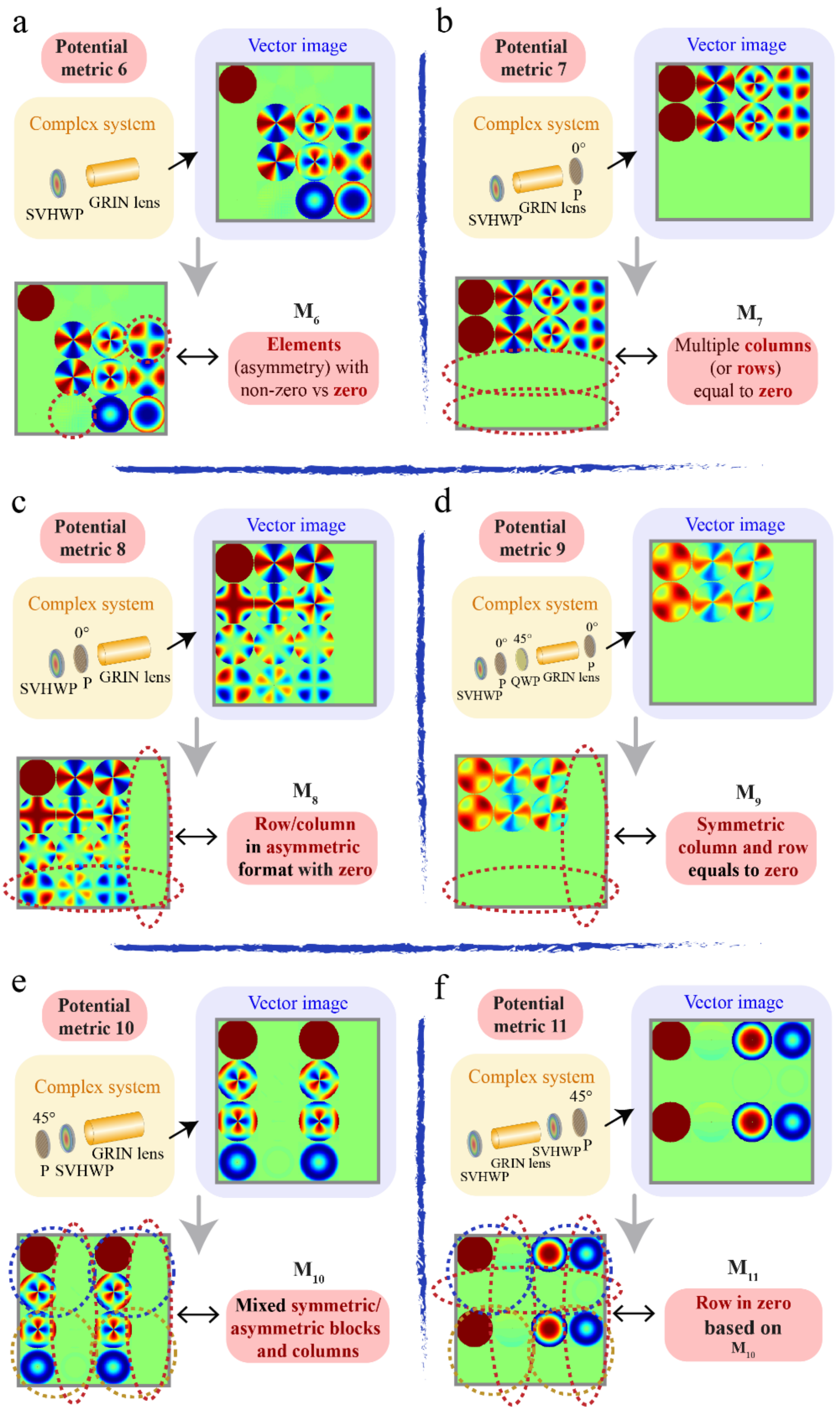

**Supplementary Figure 12. Six additional proposed metrics.** For metrics 6 to 11, they are illustrated using MMs of different GRIN lens cascades. The metrics belong to the different types mentioned in Supplementary Figure 11. All MMs are normalized as a ratio to $m_{11}$.

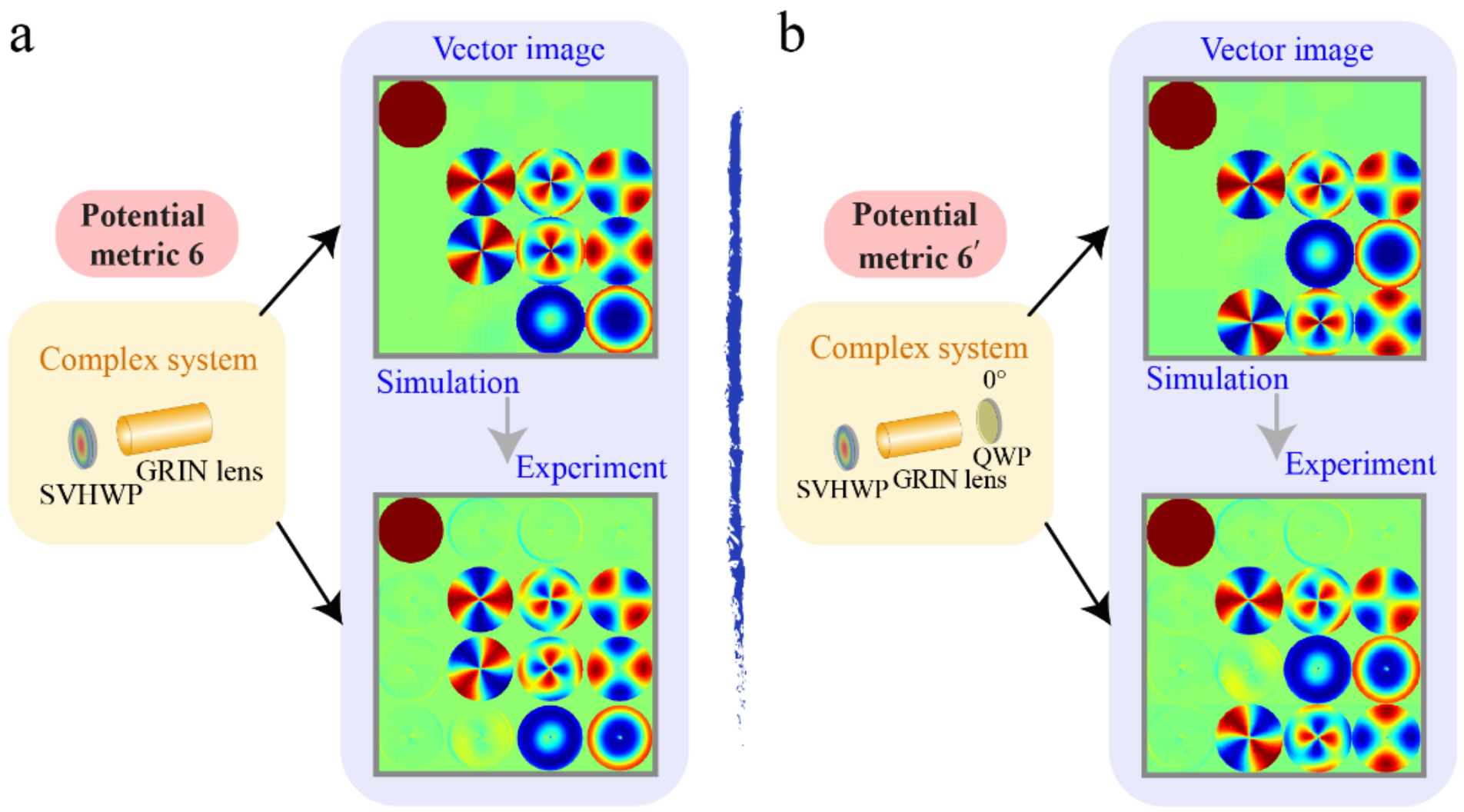

**Supplementary Figure 13. Validations between simulated and experimental MM data.** Sketches of (a) proposed metric 6 and (b) its modified format (proposed metric 6'; which belongs to "ratio metric" with zero-valued one element) show the corresponding GRIN lens cascades. Simulations and experimental MM data are presented.

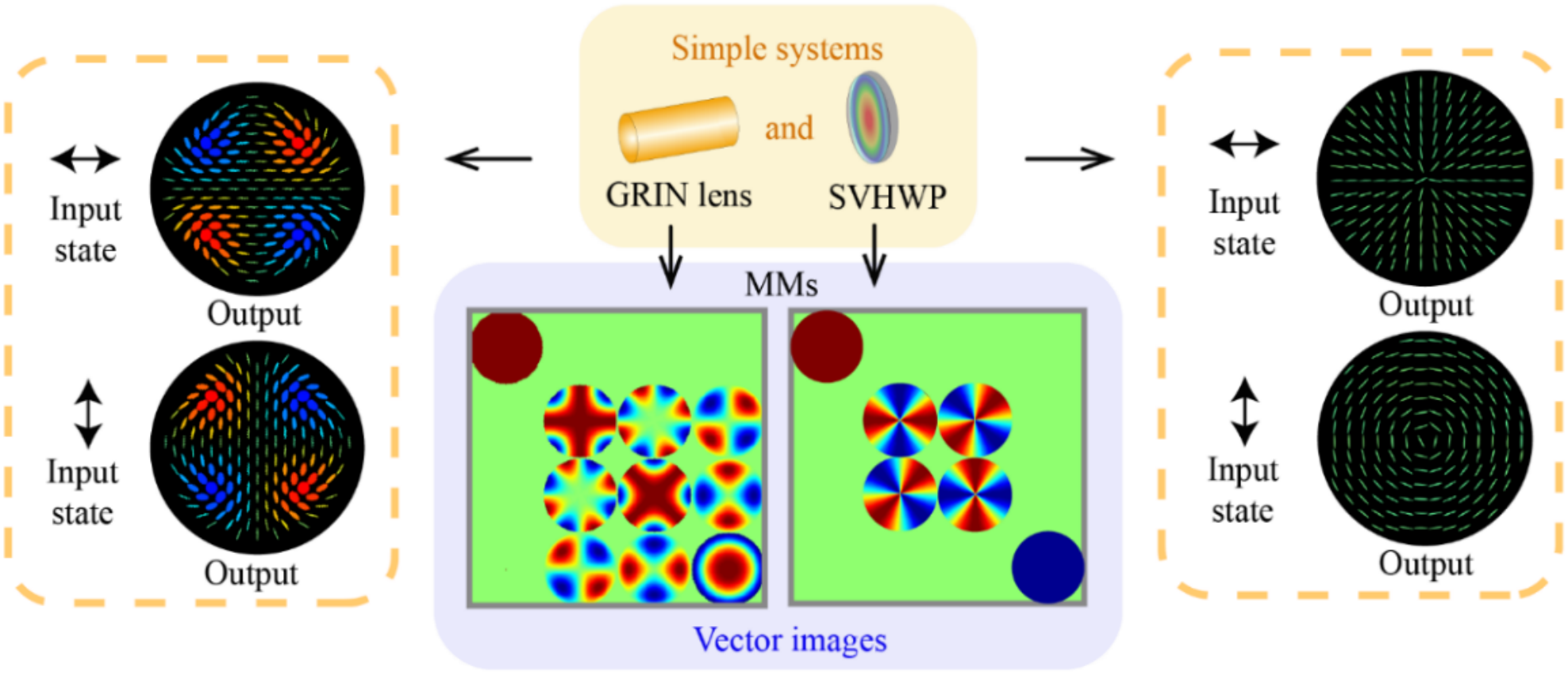

**Supplementary Figure 14. MMs and generated vector beams of GRIN lens and SVHWP.** MMs and vector beams generated via GRIN lens and SVHWP are given, the input polarisation states are also shown in the figure.

## References

1. Goldstein, D. H. *Polarized Light*. (CRC Press, 2017).

2. Chipman, R. A., Lam, W.-S. T. & Young, G. *Polarized light and optical systems*. (CRC Press, 2018).

3. Pérez, J. J. G. & Ossikovski, R. *Polarized light and the Mueller matrix approach*. (CRC Press, 2017).

4. Lu, S.-Y. & Chipman, R. A. Interpretation of Mueller matrices based on polar decomposition. *J. Opt. Soc. Am. A* **13**, 1106–1113 (1996).

5. He, H. *et al.* A possible quantitative Mueller matrix transformation technique for anisotropic scattering media/Eine mögliche quantitative Müller-Matrix-Transformations-Technik für anisotrope streuende Medien. *Photonics Lasers Med.* **2**, 129–137 (2013).

6. Arteaga, O., Garcia-Caurel, E. & Ossikovski, R. Anisotropy coefficients of a Mueller matrix. *J. Opt. Soc. Am. A* **28**, 548–553 (2011).

7. Li, P., Lv, D., He, H. & Ma, H. Separating azimuthal orientation dependence in polarization measurements of anisotropic media. *Opt. Express* **26**, 3791–3800 (2018).

8. Li, P., Tariq, A., He, H. & Ma, H. Characteristic Mueller matrices for direct assessment of the breaking of symmetries. *Opt. Lett.* **45**, 706–709 (2020).

9. Chipman, R. A. Depolarization index and the average degree of polarization. *Appl. Opt.* **44**, 2490–2495 (2005).

10. He, H. *et al.* Mueller Matrix Polarimetry—An Emerging New Tool for Characterizing the Microstructural Feature of Complex Biological Specimen. *J. Light. Technol.* **37**, 2534–2548 (2019).

11. Ghosh, N. Tissue polarimetry: concepts, challenges, applications, and outlook. *J. Biomed. Opt.* **16**, 110801 (2011).

12. Tuchin, V. V. Polarized light interaction with tissues. *J. Biomed. Opt.* **21**, 071114 (2016).

13. Ramella-Roman, J. C., Saytashev, I. & Piccini, M. A review of polarization-based imaging technologies for clinical and preclinical applications. *J. Opt.* **22**, 123001 (2020).

14. Qi, J. & Elson, D. S. Mueller polarimetric imaging for surgical and diagnostic applications: a review. *J. Biophotonics* **10**, 950–982 (2017).

15. Azzam, R. M. A. Photopolarimetric measurement of the Mueller matrix by Fourier analysis of a single detected signal. *Opt. Lett.* **2**, 148–150 (1978).

16. Collett, E. Measurement of the four Stokes polarization parameters with a single circular polarizer. *Opt. Commun.* **52**, 77–80 (1984).

17. He, C. *et al.* Linear polarization optimized Stokes polarimeter based on four-quadrant detector. *Appl. Opt.* **54**, 4458–4463 (2015).

18. Chang, J., Zeng, N., He, H., He, Y. & Ma, H. Single-shot spatially modulated Stokes polarimeter based on a GRIN lens. *Opt. Lett.* **39**, 2656–2659 (2014).

19. Chang, J. *et al.* Optimization of GRIN lens Stokes polarimeter. *Appl. Opt.* **54**, 7424–7432 (2015).

20. Chang, J. *et al.* Division of focal plane polarimeter-based 3 × 4 Mueller matrix microscope: a potential tool for quick diagnosis of human carcinoma tissues. *J. Biomed. Opt.* **21**, 056002 (2016).

21. Samlan, C. T. & Viswanathan, N. K. Field-controllable Spin-Hall Effect of Light in Optical Crystals: A Conoscopic Mueller Matrix Analysis. *Sci. Rep.* **8**, 2002 (2018).

22. Bliokh, K. Y., Rodríguez-Fortuño, F. J., Nori, F. & Zayats, A. V. Spin–orbit interactions of light. *Nat. Photonics* **9**, 796–808 (2015).

23. Bliokh, K. Y. *et al.* Spin-Hall effect and circular birefringence of a uniaxial crystal plate. *Optica* **3**, 1039–1047 (2016).

24. He, C. *et al.* Complex vectorial optics through gradient index lens cascades. *Nat. Commun.* **10**, 4264 (2019).

25. Hosten, O. & Kwiat, P. Observation of the Spin Hall Effect of Light via Weak Measurements. *Science (80-. ).* **319**, 787–790 (2008).

26. Dong, Y. *et al.* Quantitatively characterizing the microstructural features of breast ductal carcinoma tissues in different progression stages by Mueller matrix microscope. *Biomed. Opt. Express* **8**, 3643–3655 (2017).

27. Salter, P. S. & Booth, M. J. Adaptive optics in laser processing. *Light Sci. Appl.* **8**, 110 (2019).

28. Guan, J., Liu, X., Salter, P. S. & Booth, M. J. Hybrid laser written waveguides in fused silica for low loss and polarization independence. *Opt. Express* **25**, 4845–4859 (2017).

29. Salter, P. S., Simmonds, R. D. & Booth, M. J. Adaptive control of pulse front tilt, the quill effect, and directional ultrafast laser writing. in (eds. Heisterkamp, A., Herman, P. R., Meunier, M. & Nolte, S.) 861111 (2013). doi:10.1117/12.2005082.

30. Guan, J., Liu, X. & Booth, M. J. Ultrafast laser writing quill effect in low loss waveguide fabrication regime. *Opt. Express* **26**, 30716–30723 (2018).

31. Bhardwaj, V. R. *et al.* Stress in femtosecond-laser-written waveguides in fused silica. *Opt. Lett.* **29**, 1312–1314 (2004).

32. Fernandes, L. A., Grenier, J. R., Herman, P. R., Aitchison, J. S. & Marques, P. V. S. Stress induced birefringence tuning in femtosecond laser fabricated waveguides in fused silica. *Opt. Express* **20**, 24103–24114 (2012).

33. Shimotsuma, Y., Kazansky, P. G., Qiu, J. & Hirao, K. Self-Organized Nanogratings in Glass Irradiated by Ultrashort Light Pulses. *Phys. Rev. Lett.* **91**, 247405 (2003).

34. Hirao, K. & Miura, K. Writing waveguides and gratings in silica and related materials by a femtosecond laser. *J. Non. Cryst. Solids* **239**, 91–95 (1998).

35. Hnatovsky, C. *et al.* Pulse duration dependence of femtosecond-laser-fabricated nanogratings in fused silica. *Appl. Phys. Lett.* **87**, (2005).

36. Richter, S. *et al.* Nanogratings in fused silica: Formation, control, and applications. *J. Laser Appl.* **24**, 042008 (2012).

37. Cavillon, M., Wang, Y., Poumellec, B., Brisset, F. & Lancry, M. Erasure of nanopores in silicate glasses induced by femtosecond laser irradiation in the Type II regime. *Appl. Phys. A* **126**, 876 (2020).

38. Bellouard, Y. *et al.* Stress-state manipulation in fused silica via femtosecond laser irradiation. *Optica* **3**, 1285–1293 (2016).

39. Desmarchelier, R., Poumellec, B., Brisset, F., Mazerat, S. & Lancry, M. In the Heart of Femtosecond Laser Induced Nanogratings: From Porous Nanoplanes to Form Birefringence. *World J. Nano Sci. Eng.* **05**, 115–125 (2015).

40. Wang, L., Jacques, S. L. & Zheng, L. MCML—Monte Carlo modeling of light transport in multi-layered tissues. *Comput. Methods Programs Biomed.* **47**, 131–146 (1995).

41. Yun, T. *et al.* Monte Carlo simulation of polarized photon scattering in anisotropic media. *Opt. Express* **17**, 16590–16602 (2009).

42. Minna, J. D., Roth, J. A. & Gazdar, A. F. Focus on lung cancer. *Cancer Cell* **1**, 49–52 (2002).

43. Herbst, R. S., Morgensztern, D. & Boshoff, C. The biology and management of non-small cell lung cancer. *Nature* **553**, 446–454 (2018).

44. Reck, M., Heigener, D. F., Mok, T., Soria, J.-C. & Rabe, K. F. Management of non-small-cell lung cancer: recent developments. *Lancet* **382**, 709–719 (2013).

45. Litin M.D., S. C. *Mayo Clinic Family Health Book*. (2009).

46. Selman, M., King, T. E. & Pardo, A. Idiopathic Pulmonary Fibrosis: Prevailing and Evolving Hypotheses about Its Pathogenesis and Implications for Therapy. *Ann. Intern. Med.* **134**, 136–151 (2001).